\documentclass{thisIsNotJFP}

\usepackage{code}
\usepackage{comment}
\usepackage{url}
\usepackage{fancyvrb}
\usepackage{graphics}
\usepackage{boxedminipage}

\usepackage[bookmarks=false,%
            colorlinks,linkcolor=black,urlcolor=blue,%
            pdfauthor={Oleg and Ralf},%
            pdftitle={OOHaskell}]{hyperref}

\DefineShortVerb{\|}

\newcommand{\omitnow}[1]{}

\newcommand{\noskip}{\topsep0pt \parskip0pt \partopsep0pt}
\newcommand{\w}[1]{\textit{#1}}
\newcommand{\cmt}[1]{\mbox{\textrm{\emph{#1}}}}
\newcommand{\mysize}{\small}

\newcommand{\myinput}[1]{The content of this section is not yet released.}
\newcommand{\HList}{\textsc{HList}}
\newcommand{\OOHaskell}{\textsc{OOHaskell}}
\newcommand{\undefined}{\ensuremath{\bot}}

\begin{document}

 
\title{Haskell's overlooked object system\\
{\normalsize ---~\today~---}}

\author[O. Kiselyov and R. L{\"a}mmel]{Oleg Kiselyov\\
Fleet Numerical Meteorology and Oceanography Center, Monterey, CA
\and
Ralf L{\"a}mmel\\Microsoft Corp., Redmond, WA}

\maketitle

\begin{abstract}

Haskell provides type-class-bounded and parametric polymorphism
as opposed to subtype polymorphism of object-oriented languages such
as Java and OCaml. It is a contentious question whether Haskell~98
without extensions, or with common extensions, or with new extensions
can fully support conventional object-oriented programming with
encapsulation, mutable state, inheritance, overriding, statically
checked implicit and explicit subtyping, and so on.

\smallskip

In a first phase, we demonstrate how far we can get with
object-oriented functional programming, if we restrict ourselves to
plain Haskell~98. In the second and major phase, we systematically
substantiate that Haskell~98, with some common extensions, supports
all the conventional OO features plus more advanced ones, including
first-class lexically scoped classes, implicitly polymorphic classes,
flexible multiple inheritance, safe downcasts and safe co-variant
arguments. Haskell indeed can support width and depth, structural and
nominal subtyping.  We address the particular challenge to preserve
Haskell's type inference even for objects and object-operating
functions. Advanced type inference is a strength of Haskell that is
worth preserving. Many of the features we get ``for free'': the type
system of Haskell turns out to be a great help and a guide rather than
a hindrance.

\smallskip

The OO features are introduced in Haskell as the \OOHaskell\ 
\emph{library}, non-trivially based on the \HList\ library of
extensible polymorphic records with first-class labels and
subtyping. The library sample code, which is patterned after the
examples found in OO textbooks and programming language tutorials,
including the OCaml object tutorial, demonstrates that OO code translates
into \OOHaskell\ in an intuition-preserving way: essentially
expression-by-expression, without requiring global transformations.

\smallskip

\OOHaskell\ lends itself as a sandbox for typed OO language design.

\medskip


\end{abstract}

\makeatactive

\newpage 

{

\footnotesize

\setcounter{tocdepth}{3}
\tableofcontents

}

\newpage


\section{Introduction}

{ \sloppypar

The topic of object-oriented programming in the functional language
Haskell is raised time and again on programming language mailing
lists, on programming tutorial websites, and in verbal communication
at programming language conferences with remarkable
intensity. Dedicated OO Haskell language extensions have been
proposed; specific OO idioms have been encoded in
Haskell~\cite{HS95,GJ96,FLMPJ99, SPJ01,Nordlander02,MonadReader3}.
The interest in this topic is not at all restricted to Haskell
researchers and practitioners since there is a fundamental and unsettled
question~---~a question that is addressed in the present paper:\footnote{On a more anecdotal account, we have collected
informative pointers to mailing list discussions, which document the
unsettled understanding of OO programming in Haskell and the relation
between OO classes and Haskell's type classes: {\scriptsize
\url{http://www.cs.mu.oz.au/research/mercury/mailing-lists/mercury-users/mercury-users.0105/0051.html},
\url{http://www.talkaboutprogramming.com/group/comp.lang.functional/messages/47728.html},
\url{http://www.haskell.org/pipermail/haskell/2003-December/013238.html},
\url{http://www.haskell.org/pipermail/haskell-cafe/2004-June/006207.html},
\url{http://www.haskell.org//pipermail/haskell/2004-June/014164.html}}}

}

\begin{center}\itshape
What is the relation between type-class-bounded and subtype polymorphism?
\end{center}

\noindent
In this research context, we specifically (and emphatically) restrict
ourselves to the existing Haskell language (Haskell~98 and common
extensions where necessary), i.e., no new Haskell extensions are to be
proposed. As we will substantiate, this restriction is adequate, as it
allows us to deliver a meaningful and momentous answer to the
aforementioned question. At a more detailed level, we offer the
following motivation for research on OO programming in Haskell:
\begin{itemize}

\item
In an \emph{intellectual} sense, one may wonder whether Haskell's
advanced type system is expressive enough to model object types,
inheritance, subtyping, virtual methods, etc. No general, conclusive
result has been available so far.

\smallskip

\item
In a \emph{practical} sense, one may wonder whether we can faithfully
transport imperative OO designs from, say, C\#, C++, Eiffel, Java, VB
to Haskell~---~without totally rewriting the design and without
foreign-language interfacing.

\smallskip

\item From a \emph{language design} perspective, Haskell has a strong
record in prototyping semantics and encoding abstraction mechanisms,
but one may wonder whether Haskell can perhaps even serve as a sandbox
for design of typed object-oriented languages so that one can play
with new ideas without the immediate need to write or modify a
compiler.

\smallskip

\item In an \emph{educational} sense, one may wonder whether
more or less advanced functional and object-oriented programmers can
improve their understanding of Haskell's type system and OO concepts
by looking into the pros and cons of different OO encoding options in
Haskell.

\end{itemize}

\smallskip

This paper delivers substantiated, positive answers to these
questions. We describe \OOHaskell~---~a Haskell-based library for (as
of today: imperative) OO programming in Haskell. \OOHaskell\ delivers
Haskell's ``overlooked'' object system. The key to this result is a
good deal of exploitation of Haskell's advanced type system
\emph{combined} with a careful identification of a suitable object
encoding. We instantiate and enhance existing encoding
techniques (such as~\cite{PT94,ML-ART,AC96}) aiming
at a practical object system that blends well with the
host language~---~Haskell. We take advantage of our previous work on
heterogeneous collections~\cite{HLIST-HW04} (the \HList\ library).
More generally, we put type-class-based or type-level programming
to work~\cite{Hallgren01,Fake,NTGS02,NTGS01}.

\medskip

\noindent
The simplified story is the following:

\noindent
- Classes are represented as functions that are in fact object generators.

\noindent
- State is maintained through mutable variables allocated by object generators.

\noindent
- Objects are represented as records of closures with a component for
each method.

\noindent
- Methods are monadic functions that can access state and |self|.

\noindent
- We use \HList's record calculus (extensible records, up-casts, etc.).

\noindent
- We use type-class-based functionality to program the object typing rules.

\medskip

\noindent
To deliver a faithful, convenient and comprehensive object system,
several techniques had to be discovered and combined. Proper effort
was needed to preserve Haskell's type inference for OO programming
idioms (as opposed to explicit type declarations or type constraints
for classes, methods, and up-casts). The obtained result,
\OOHaskell, delivers an amount of polymorphism and type inference that
is unprecedented. Proper effort was also needed in order to deploy
value recursion for closing object generators. Achieving safety of
this approach was a known challenge \cite{ML-ART}. In order to fully
appreciate the object system of \OOHaskell, we also review less
sophisticated, less favourable encoding alternatives.

{\sloppypar

Not only \OOHaskell\ provides the conventional OO idioms; we have also
language-engineered several features that are either bleeding-edge or
unattainable in mainstream OO languages: for example, first-class
classes and class closures; statically type-checked collection classes
with bounded polymorphism of implicit collection arguments; multiple
inheritance with user-controlled sharing; safe co-variant argument
subtyping. It is remarkable that these and more familiar
object-oriented features are not introduced by fiat~---~we get them
for free. For example, the type of a collection with bounded
polymorphism of elements is inferred automatically by the
compiler. Also, abstract classes are uninstantiatable not because we
say so but because the program will not typecheck otherwise. Co- and
contra-variant subtyping rules and the safety conditions for the
co-variant method argument types are checked automatically without any
programming on our part. These facts suggest that (OO)Haskell lends
itself as prime environment for typed object-oriented language design.

}


\newpage

\subsection*{Road-map of this paper}

\begin{itemize}
\item Sec.~\ref{S:shapes}: We encode a tutorial OO example both in C++ and \OOHaskell.
\item Sec.~\ref{S:poor}: We review alternative object encodings in Haskell~98 and beyond.
\item Sec.~\ref{S:OOHaskell1} and Sec.~\ref{S:OOHaskell2}: We describe all \OOHaskell\ idioms.
The first part focuses on idioms where subtyping and object types do
not surface the OO program code. The second part covers all technical details of subtyping
including casts and variance properties.
\item Sec.~\ref{S:discuss}: We discuss usability issues, related work and future work.
\item Sec.~\ref{S:concl}: We conclude the paper.
\end{itemize}

\noindent
The main sections, Sec.~\ref{S:OOHaskell1} and
Sec.~\ref{S:OOHaskell2}, are written in tutorial style, as to ease
digestion of all techniques, as to encourage OO programming and OO
language design experiments. There is an extended source distribution
available.\footnote{The source code can be downloaded at
\url{http://www.cwi.nl/~ralf/OOHaskell/}, and it is subject to 
a very liberal license (MIT/X11 style). As of writing, the actual code
commits to a few specific extensions of the GHC implementation of
Haskell~---~for reasons of convenience. In principle, Haskell~98 +
multi-parameter classes with functional dependencies is sufficient.}


\section{The folklore `shapes' example}
\label{S:shapes}

One of the main goals of this paper is to be able to represent the
conventional OO code, in as straightforward way as possible.  The
implementation of our system may be not for the feeble at
heart~---~however, the user of the system must be able to write
conventional OO code without understanding the complexity of the
implementation. Throughout the paper, we illustrate \OOHaskell\ with a
series of practical examples as they are commonly found in OO
textbooks and programming language tutorials. In this section, we
begin with the so-called `shapes' example.

\begin{figure}[t!]
\bigskip
\bigskip
\begin{center}
\resizebox{.77\textwidth}{!}{\includegraphics{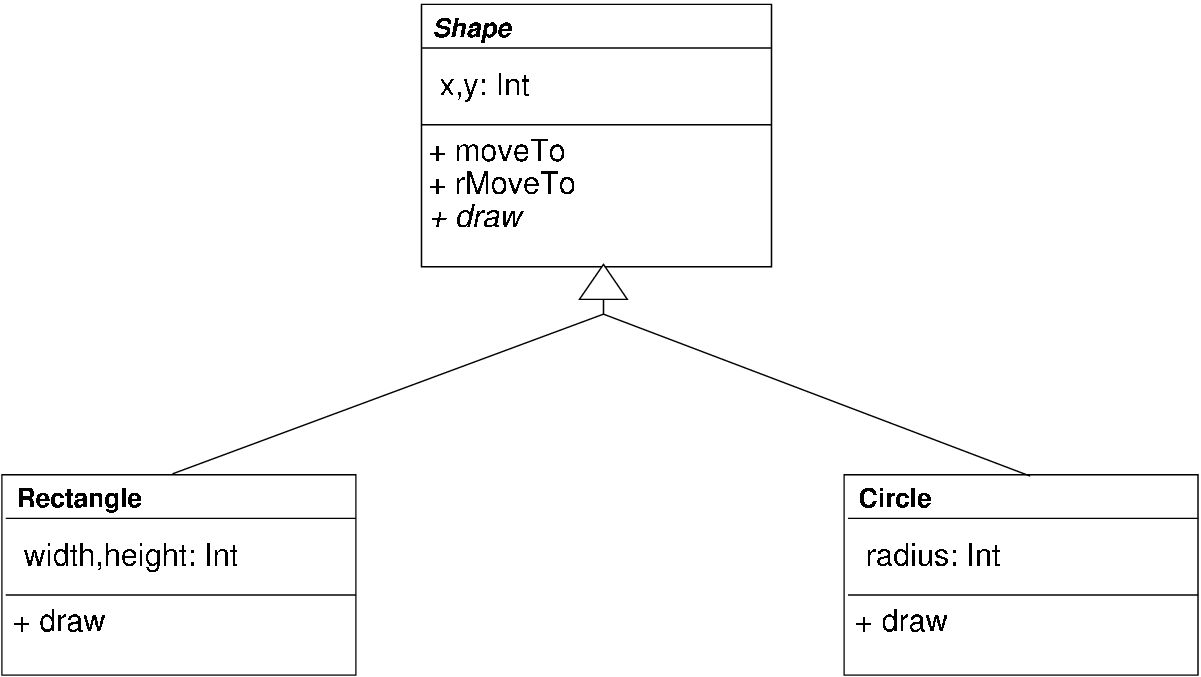}}
\end{center}
\caption{Shapes with state and a subtype-specific draw method}
\label{F:shapes}
\bigskip
\bigskip
\end{figure}

We face a type for `shapes' and two subtypes for `rectangles' and
`circles'; see Fig.~\ref{F:shapes}. Shapes maintain coordinates as
state. Shapes can be moved around and drawn. The exercise shall be to
place objects of \emph{different} kinds of shapes in a collection and
to iterate over them as to draw the shapes. It turns out that this
example is a crisp OO benchmark.\footnote{The `shapes' problem
has been designed by Jim Weirich and deeply explored by him and Chris
Rathman. See the multi-lingual collection `OO Example Code' by Jim
Weirich at \url{http://onestepback.org/articles/poly/}; see also an
even heavier collection `OO Shape Examples' by Chris Rathman at
\url{http://www.angelfire.com/tx4/cus/shapes/}.}


\subsection{C++ reference encoding}

The type of shapes can be defined as a C++ class as follows:

\begin{Verbatim}[fontsize=\small,commandchars=\|\!\@]
 class Shape {
 public:
\end{Verbatim}

\begin{Verbatim}[fontsize=\small,commandchars=\|\!\@]
   // |cmt!Constructor method@
   Shape(int newx, int newy) {
     x = newx;
     y = newy;
   }
\end{Verbatim}

\begin{Verbatim}[fontsize=\small,commandchars=\|\!\@]
   // |cmt!Accessors@
   int getX() { return x; }
   int getY() { return y; }
   void setX(int newx) { x = newx; }
   void setY(int newy) { y = newy; }
\end{Verbatim}

\begin{Verbatim}[fontsize=\small,commandchars=\|\!\@]
   // |cmt!Move shape position@
   void moveTo(int newx, int newy) {
      x = newx;
      y = newy;
   }
\end{Verbatim}

\begin{Verbatim}[fontsize=\small,commandchars=\|\!\@]
   // |cmt!Move shape relatively@
   void rMoveTo(int deltax, int deltay) {
      moveTo(getX() + deltax, getY() + deltay);
   }
\end{Verbatim}

\begin{Verbatim}[fontsize=\small,commandchars=\|\!\@]
   // |cmt!An abstract draw method@
   virtual void draw() = 0;
\end{Verbatim}

\begin{Verbatim}[fontsize=\small,commandchars=\|\!\@]
 // |cmt!Private data@
 private:
   int x;
   int y;
 }
\end{Verbatim}
The @x@, @y@ coordinates are private, but they can be
accessed through getters and setters. The methods for accessing and
moving shapes are inherited by the subclasses of @Shape@. The @draw@
method is virtual and even abstract; hence concrete subclasses must
implement @draw@.

The subclass @Rectangle@ is derived as follows:

\begin{Verbatim}[fontsize=\small,commandchars=\|\!\@]
 class Rectangle: public Shape {
 public:
\end{Verbatim}

\begin{Verbatim}[fontsize=\small,commandchars=\|\!\@]
   // |cmt!Constructor method@
   Rectangle(int newx, int newy, int newwidth, int newheight)
            : Shape(newx, newy) {
     width  = newwidth;
     height = newheight;
   }
\end{Verbatim}

\begin{Verbatim}[fontsize=\small,commandchars=\|\!\@]
   // |cmt!Accessors@
   int getWidth()  { return width; }
   int getHeight() { return height; }
   void setWidth(int newwidth)   { width = newwidth; }
   void setHeight(int newheight) { height = newheight; }
\end{Verbatim}

\begin{Verbatim}[fontsize=\small,commandchars=\|\!\@]
   // |cmt!Implementation of the abstract draw method@
   void draw() {
     cout << "Drawing a Rectangle at:("
          << getX() << "," << getY()
          << "), width " << getWidth()
          << ", height " << getHeight() << endl;
   }
\end{Verbatim}

\begin{Verbatim}[fontsize=\small,commandchars=\|\!\@]
 // |cmt!Additional private data@
 private:
    int width;
    int height;
 };
\end{Verbatim}

For brevity, we elide the similar derivation of the subclass @Circle@:

\begin{code}
 class Circle : public Shape {
  Circle(int newx, int newy, int newradius)
        : Shape(newx, newy) { ... }
  ...
 }
\end{code}

The following code block constructs different shape objects and
invokes their methods. More precisely, we place two shapes of
different kinds in an array, @scribble@, and then loop over it to
draw and move the shape objects:

\begin{code}
 Shape *scribble[2];
 scribble[0] = new Rectangle(10, 20, 5, 6);
 scribble[1] = new Circle(15, 25, 8);
 for (int i = 0; i < 2; i++) {
   scribble[i]->draw();
   scribble[i]->rMoveTo(100, 100);
   scribble[i]->draw();
 }
\end{code}
The loop over @scribble@ exercises subtyping
polymorphism: the actually executed implementation of the @draw@
method differs per element in the array. The program run produces the
following output~---~due to the logging-like implementations of the
@draw@ method:

\begin{code}
 Drawing a Rectangle at:(10,20), width 5, height 6
 Drawing a Rectangle at:(110,120), width 5, height 6
 Drawing a Circle at:(15,25), radius 8
 Drawing a Circle at:(115,125), radius 8
\end{code}


\subsection{OOHaskell encoding}

We now show an \OOHaskell\ encoding, which happens to pleasantly mimic
the C++ encoding, while any remaining deviations are appreciated. Most
notably, we are going to leverage type inference: we will not define
\emph{any} type. The code shall be fully statically typed
nevertheless.

\smallskip

Here is the \OOHaskell\ rendering of the shape class:

\begin{Verbatim}[fontsize=\small,commandchars=\|\{\}]
 -- |cmt{Object generator for shapes}
 shape newx newy self
   = do
	-- |cmt{Create references for private state}
        x <- newIORef newx
        y <- newIORef newy
\end{Verbatim}

\begin{Verbatim}[fontsize=\small,commandchars=\|\{\}]
	-- |cmt{Return object as record of methods}
        returnIO $  getX     .=. readIORef x
                .*. getY     .=. readIORef y
                .*. setX     .=. writeIORef x
                .*. setY     .=. writeIORef y
	        .*. moveTo   .=. (\newx newy -> do
                                        (self # setX) newx
                                        (self # setY) newy )
                .*. rMoveTo  .=. (\deltax deltay ->
                            do
                              x  <- self # getX
                              y  <- self # getY
                              (self # moveTo) (x + deltax) (y + deltay) )
                .*. emptyRecord
\end{Verbatim}


Classes become \emph{functions that take constructor arguments plus a
self reference and that return a computation whose result is the new
object~---~a record of methods} including getters and setters. We can
invoke methods of the same object through @self@; cf.\ the method
invocation @self@~@#@~@getX@ and others. (The infix operator
@#@ denotes method invocation.) Our objects are mutable, implemented
via @IORef@. (@STRef@ also suffices.) Since most OO systems in practical
use have mutable state, \OOHaskell\ does not (yet) offer
\emph{functional objects}, which are known to be challenging on their
own. We defer functional objects to future work.

We use the extensible records of the \HList\ library~\cite{HLIST-HW04},
hence:
\begin{itemize}\noskip
\item @emptyRecord@ denotes what the name promises,
\item @(.*.)@ stands for (right-associative) record extension,
\item @(.=.)@ is record-component construction: \w{label} @.=.@ \w{value},
\item Labels are defined according to a trivial scheme, to be explained later.
\end{itemize}
The abstract @draw@ method is not mentioned in the
\OOHaskell\ code because it is not used in any other method, neither did
we dare declaring its type. As a side effect, the object generator
@shape@ is instantiatable whereas the explicit declaration of the
abstract @draw@ method made the C++ class @Shape@ uninstantiatable. We
will later show how to add similar declarations in \OOHaskell.

\medskip

\noindent
We continue with the \OOHaskell\ code for the shapes example.

\begin{Verbatim}[fontsize=\small,commandchars=\|\{\}]
 -- |cmt{Object generator for rectangles}
 rectangle newx newy width height self
   = do
        -- |cmt{Invoke object generator of superclass}
        super <- shape newx newy self
\end{Verbatim}

\begin{Verbatim}[fontsize=\small,commandchars=\|\{\}]
        -- |cmt{Create references for extended state}
        w <- newIORef width
        h <- newIORef height
\end{Verbatim}

\begin{Verbatim}[fontsize=\small,commandchars=\|\{\}]
	-- |cmt{Return object as record of methods}
        returnIO $
            getWidth  .=. readIORef w
        .*. getHeight .=. readIORef h
        .*. setWidth  .=. (\neww -> writeIORef w neww)
        .*. setHeight .=. (\newh -> writeIORef h newh)
\end{Verbatim}


\begin{Verbatim}[fontsize=\small,commandchars=\|\{\}]
        .*. draw .=. 
             do -- |cmt{Implementation of the abstract draw method}
                putStr  "Drawing a Rectangle at:(" <<
                        self # getX << ls "," << self # getY <<
                        ls "), width " << self # getWidth <<
                        ls ", height " << self # getHeight << ls "\n"
\end{Verbatim}

\begin{Verbatim}[fontsize=\small,commandchars=\|\{\}]
            -- |cmt{Rectangle records start from shape records}
        .*. super
\end{Verbatim}
This snippet illustrates the essence of inheritance in \OOHaskell.
Object generation for the supertype is made part of the monadic
sequence that defines object generation for the subtype; @self@ is
passed from the subtype to the supertype. Subtype records are derived
from supertype records through record extension (or potentially also
through record updates when overrides are to be modelled).

As in the C++ case, we elide the derivation of the object generators
for circles:

\begin{code}
 circle newx newy newradius self
  = do
       super <- shape newx newy self
       ...
       returnIO ... .*. super
\end{code}

Ultimately, here is the \OOHaskell\ rendering of the `scribble loop':

\begin{Verbatim}[fontsize=\small,commandchars=\|\{\}]
 -- |cmt{Object construction and invocation as a monadic sequence}
 myOOP = do
\end{Verbatim}

\begin{Verbatim}[fontsize=\small,commandchars=\|\{\}]
          -- |cmt{Construct objects}
          s1 <- mfix (rectangle (10::Int) (20::Int) 5 6)
          s2 <- mfix (circle (15::Int) 25 8)
\end{Verbatim}

\begin{Verbatim}[fontsize=\small,commandchars=\|\{\}]
          -- |cmt{Create a homogeneous list of different shapes}
          let scribble = consLub s1 (consLub s2 nilLub)
\end{Verbatim}

\begin{Verbatim}[fontsize=\small,commandchars=\|\{\}]
          -- |cmt{Loop over list with normal monadic map}
          mapM_ (\shape -> do
                            shape # draw
                            (shape # rMoveTo) 100 100
                            shape # draw)
                scribble
\end{Verbatim}

The use of @mfix@ (an analogue of the @new@ in C++) reflects that object
generators take `self' and construct (part of) it. Open recursion
enables inheritance. The @let@~@scribble@~\ldots binding is noteworthy.
We cannot \emph{directly} place rectangles and circles in a normal
Haskell list~---~the following cannot possibly type check:

\begin{Verbatim}[fontsize=\small,commandchars=\|\{\}]
          let scribble = [s1,s2] -- s1 and s2 are of different types!
\end{Verbatim}

\noindent
We have to homogenise the types of @s1@ and @s2@ when
forming a Haskell list. To this end, we use special list constructors
@nilLub@ and @consLub@ as opposed to the normal list constructors @[]@
and @(:)@. These new constructors coerce the list elements to the
least-upper bound type of all the element types. Incidentally, if the
`intersection' of the types of the objects @s1@ and @s2@ does not
include the methods that are invoked later (i.e., @draw@ and @rMoveTo@), we get a static type
error which literally says so. As a result, the original for-loop can
be carried out in the native Haskell way: a normal (monadic) list map
over a normal Haskell list of shapes. Hence, we have exercised a
faithful model of subtype polymorphism, which also allows for (almost)
implicit subtyping. \OOHaskell\ provides several subtyping models, as we
will study later.


\subsection{Discussion of the example}

\subsubsection{Classes vs.\ interfaces}

The C++ code should not be misunderstood to suggest that \emph{class}
inheritance is the only OO design option for the shapes hierarchy. In
a Java-like language, one may want to model @Shape@ as an
\emph{interface}, say, @IShape@, with @Rectangle@ and @Circle@ as classes
implementing this interface. This design would not allow us to reuse
the implementations of the accessors and the move methods. So one may
want to combine interface polymorphism \emph{and} class
inheritance. That is, the classes @Rectangle@ and @Circle@ will be
rooted by an additional implementation class for shapes, say @Shape@,
which hosts implementations shared among different shape
classes~---~incidentally a part of the @IShape@ interface. The
remainder of the @IShape@ interface, namely the @draw@ method in our
example, would be implemented in @Rectangle@ and @Circle@.

More generally, OO designs that employ interface polymorphism alone
are rare, so we need to provide encodings for both OO interface
polymorphism \emph{and} OO class inheritance in \OOHaskell. One may say
that the former mechanism is essentially covered by Haskell's type
classes (modulo the fact that we would still need an object
encoding). The latter mechanism is specifically covered by original
\HList\ and \OOHaskell\ contributions: structural subtyping polymorphism
for object types, based on polymorphic extensible records and programmable
subtyping constraints. (Sec.~\ref{S:nominal} discusses \emph{nominal}
object types in \OOHaskell\.)


\subsubsection{Extensibility and encapsulation}
\label{S:ext}

Both the C++ encoding and the \OOHaskell\ encoding of the shapes
example are faithful to the encapsulation premise as well as the
\emph{extensibility premise} of the OO paradigm.  An object
encapsulates \emph{both} data (`state') and methods (`behaviour').  One
may add new kinds of shapes without rewriting (or, perhaps, even
re-compiling) existing code.

Both premises are the subject of an unsettled debate in the
programming language community, especially with regards to functional
programming. The basic OO paradigm has been criticised~\cite{ZO04} for
its over-emphasis of extensibility in the subtyping dimension and for its
neglect of other dimensions such as the addition of new functions into
a pre-existing subtyping hierarchy. While we agree with this overall
criticism, we avoid the debate in this paper. We simply want
\OOHaskell\ to provide an object encoding that is compatible with the
established OO paradigm. (Incidentally, some of the
non-encapsulation-based encodings in Sec.~\ref{S:poor} show that
Haskell supports extensibility in both the data and the functionality
dimension.)


\subsubsection{Subtyping technicalities}

The ``scribble loop'' is by no means a contrived scenario. It is a
faithful instance of the ubiquitous composite design
pattern~\cite{GHJV94}. In terms of expressiveness and typing
challenges, this sort of loop over an array of shapes of different
kinds forces us to explore the tension between implicit and explicit
subtyping. As we will discuss, it is relatively straightforward to use
type-class-bounded polymorphism to represent subtype constraints. It
is however less straightforward to accumulate entities of different
subtypes in the same collection. With explicit subtyping (e.g., by
wrapping in a properly constrained existential envelope) the burden
would be on the side of the programmer. A key challenge for \OOHaskell\ 
was to make subtyping (almost) implicit in all the cases, where a OO
programmer would expect it. This is a particular area in which
\OOHaskell\ goes beyond OCaml~\cite{OCaml}---~the de-facto leading
strongly typed functional object-oriented language. \OOHaskell\ provides
a range of subtyping notions, including one that even allows for safe
downcasts for object types. This is again something that has not been
achieved in OCaml to date.


\section{Alternative Haskell encodings}
\label{S:poor}

\OOHaskell\ goes particularly far in providing an object system, when
compared to conservative Haskell programming knowledge. To this end,
we put type-class-based or type-level programming to work. In this
section, we will review more conservative object encodings with their
characteristics and limitations. All of them require boilerplate code
from the programmer.

Some of the `conservative' encodings to come are nevertheless involved
and enlightening. In fact, the full spectrum of encodings has not been
documented before~---~certainly not in a Haskell context. So we reckon
that their detailed analysis makes a useful contribution. Furthermore,
several of the discussed techniques are actually used in \OOHaskell,
where some of them are simply generalised through the advanced use of
Haskell's type classes. Hence, the present section is an incremental preparation
for the main sections Sec.~\ref{S:OOHaskell1} and Sec.~\ref{S:OOHaskell2}.

For most of this section, we limit ourselves to Haskell~98. (By
contrast, \OOHaskell\ requires several common Haskell~98 extensions.) 
Towards the end of the section, we will investigate the value of
dismissing this restriction.


\subsection{Map subtype hierarchy to an algebraic datatype}
\label{S:lennart}

We begin with a trivial and concise encoding. Its distinguishing
characteristic is extreme simplicity.\footnote{Thanks to Lennart
  Augustsson for pointing out this line of encoding.\\ Cf.\
  \url{http://www.haskell.org/pipermail/haskell/2005-June/016061.html}}
It uses only basic Haskell~98 idioms.  The encoding is also seriously
limited, lacking extensibility with regard to new forms of shapes
(cf.\ Sec.~\ref{S:ext}).

We define an algebraic datatype for shapes, where each kind of shape
amounts to a constructor declaration. For readability, we use labelled
fields instead of unlabelled constructor components.

\begin{code}
 data Shape =
              Rectangle { getX      :: Int
                        , getY      :: Int 
                        , getWidth  :: Int 
                        , getHeight :: Int }
            |
              Circle { getX      :: Int
                     , getY      :: Int 
                     , getRadius :: Int }
\end{code}

Both constructor declarations involve labelled fields for the $(x,y)$
position of a shape. While this reusability dimension is not
emphasised at the datatype level, we can easily define reusable
setters for the position.  (There are some issues regarding type
safety, which we will address later.) For instance:

\begin{code}
 setX :: Int -> Shape -> Shape
 setX i s = s { getX = i } 
\end{code}
We can also define setters for @Rectangle@- and @Circle@-specific
fields. For instance:

\begin{code}
 setWidth :: Int -> Shape -> Shape
 setWidth i s = s { getWidth = i } 
\end{code}
It is also straightforward to define functions for moving around shapes:

\begin{code}
 moveTo :: Int -> Int -> Shape -> Shape
 moveTo x y = setY y . setX x 
\end{code}

\begin{code}
 rMoveTo :: Int -> Int -> Shape -> Shape
 rMoveTo deltax deltay s = moveTo x y s
  where
   x = getX s + deltax
   y = getY s + deltay
\end{code}

The function for drawing shapes properly discriminates on the kind of
shapes. That is, there is one equation per kind of shape. Subtype polymorphism
reduces to pattern matching, so to say:

\begin{code}
 draw :: Shape -> IO ()
\end{code}

\begin{code}
 draw s@(Rectangle _ _ _ _)
          =  putStrLn ("Drawing a Rectangle at:("
          ++ (show (getX s))
          ++ ","
          ++ (show (getY s))
          ++ "), width " ++ (show (getWidth s))
          ++ ", height " ++ (show (getHeight s)))
\end{code}

\begin{code}
 draw s@(Circle _ _ _)
          =  putStrLn ("Drawing a Circle at:("
          ++ (show (getX s))
          ++ ","
          ++ (show (getY s))
          ++ "), radius "
          ++ (show (getRadius s)))
\end{code}
With this encoding, it is trivial to build a collection of shapes of different
kinds and to iterate over it such that each shape is drawn and moved (and drawn again):

\begin{code}
 main =
       do
          let scribble = [ Rectangle 10 20 5 6
                         , Circle 15 25 8
                         ]
          mapM_ ( \x -> 
                    do
                       draw x
                       draw (rMoveTo 100 100 x))
                scribble
\end{code}

\paragraph{Assessment of the encoding}

\mbox{}

\begin{itemize}

\item
The encoding ignores the encapsulation premise of the OO paradigm.

\smallskip

\item
The foremost weakness of the encoding is the lack of extensibility.
The addition of a new kind of shape would require re-compilation of
all code; it would also require amendments of existing definitions or
declarations: the datatype declaration @Shape@ and the function
definition @draw@.

\smallskip

\item
A related weakness is that the overall scheme does not suggest a way
of dealing with virtual methods: introduce the type of a method for a
base type potentially with an implementation; define or override the
method for a subtype. We would need a scheme that offers
(explicit and implicit) open recursion for datatypes and functions
defined on them.

\smallskip

\item
The setters @setX@ and @setY@ \emph{happen} to be total because all
constructors end up defining labelled fields @getX@ and @getY@. The
type system does \emph{not} prevent us from forgetting those labels
for some constructor. It is relatively easy to resolve this issue to
the slight disadvantage of conciseness. (For instance, we may
avoid labelling entirely, and use pattern matching instead. We may
also compose together rectangles and circles from common shape data
and deltas.)

\smallskip

\item
The use of a single algebraic datatype @Shape@ implies that
@Rectangle@- and @Circle@-specific functions cannot be defined as
total functions. Such biased functions, e.g., @setWidth@, are only
defined for certain constructors. Once we go beyond the simple-minded encoding model
of this section, it will be possible to increase type
safety by making type distinctions for different kinds of shapes, but
then we will also encounter the challenge of subtype polymorphism.

\end{itemize}


\subsection{Map object data to tail-polymorphic record types}
\label{S:burton}

There is a folklore technique for encoding extensible
records~\cite{Burton90} that we can use to model the shapes hierarchy
in Haskell~98. Simple type classes let us implement virtual
methods. We meet the remaining challenge of placing different shapes
into one list by making different subtypes homogeneous through
embedding shape subtypes into a union type (Haskell's @Either@).

We begin with a datatype for extensible shapes; cf.\ @shapeTail@:

\begin{code}
 data Shape w =
      Shape { getX :: Int
            , getY :: Int
            , shapeTail :: w }
\end{code}
For convenience, we also provide a constructor for shapes:

\begin{code}
 shape x y w = Shape { getX = x
                     , getY = y
                     , shapeTail = w }
\end{code}

We can define setters and movers once and for all for all possible
extensions of @Shape@ by simply \emph{leaving the extension type
parametric}. The actual equations are literally the same as in the
previous section; so we only show the (different) parametrically
polymorphic types:

\begin{code}
 setX :: Int -> Shape w -> Shape w
 setY :: Int -> Shape w -> Shape w
 moveTo :: Int -> Int -> Shape w -> Shape w
 rMoveTo :: Int -> Int -> Shape w -> Shape w
\end{code}
The presence of the type variable @w@ expresses that the earlier
definitions on @Shape@~\ldots\ can clearly be instantiated to all
subtypes of @Shape@. The @draw@ function must be placed in a
dedicated type class, @Draw@, because we anticipate the need to
provide type-specific implementations of @draw@. (One may compare this
style with C++ where one explicitly declares a method to be
(\emph{pure}) \emph{virtual}.)

\begin{code}
 class Draw w
  where
   draw :: Shape w -> IO ()
\end{code}

Shape extensions for rectangles and circles are built according to a
common scheme. We only show the details for rectangles.  We begin with
the definition of the ``data delta'' contributed by rectangles; each
such delta is again polymorphic in its tail.

\begin{code}
 data RectangleDelta w =
      RectangleDelta { getWidth      :: Int 
                     , getHeight     :: Int
                     , rectangleTail :: w }
\end{code}
We define the type of rectangles as an instance of @Shape@:

\begin{code}
 type Rectangle w = Shape (RectangleDelta w)
\end{code}

For convenience, we provide a constructor for rectangles. Here we fix
the tail of the rectangle delta to @()@. (We could still further
instantiate @Rectangle@ and define new constructors later, if
necessary.)

\begin{code}
 rectangle x y w h
  = shape x y $ RectangleDelta {
                  getWidth      = w
                , getHeight     = h
                , rectangleTail = () }
\end{code}

%
The definition of rectangle-specific setters involves nested record
manipulation:

\begin{code}
 setHeight :: Int -> Rectangle w -> Rectangle w
 setHeight i s = s { shapeTail = (shapeTail s) { getHeight = i } }
\end{code}

\begin{code}
 setWidth :: Int -> Rectangle w -> Rectangle w
 setWidth i s = s { shapeTail = (shapeTail s) { getWidth = i } }
\end{code}

The rectangle-specific @draw@ function is defined through a @Draw@ instance:

\begin{code}
 instance Draw (RectangleDelta w)
  where
   draw s
     =   putStrLn ("Drawing a Rectangle at:("
     ++ (show (getX s))
     ++ ","
     ++ (show (getY s))
     ++ "), width "
     ++ (show (getWidth (shapeTail s)))
     ++ ", height "
     ++ (show (getHeight (shapeTail s))))
\end{code}

The difficult part is the `scribble loop'. We cannot easily form a
collection of shapes of different kinds. For instance, the following
attempt will not type-check:

\begin{code}
 -- Wrong! There is no homogeneous element type.
 let scribble = [ rectangle 10 20 5 6
                , circle 15 25 8
                ]
\end{code}

There is a relatively simple technique to make rectangles and circles
homogeneous within the scope of the @scribble@ list and its clients.
We have to establish a union type for the different kinds of
shapes.\footnote{Haskell~98 supports unions in the prelude: with the
type name @Either@, and the two constructors @Left@ and @Right@ for
the branches of the union.} Using an appropriate helper, @tagShape@, for
embedding shapes into a union type (Haskell's @Either@), we may
construct a homogeneous collection as follows:

\begin{code}
 let scribble = [ tagShape Left  (rectangle 10 20 5 6)
                , tagShape Right (circle 15 25 8)
                ]
\end{code}
The boilerplate operation for embedding is trivially defined as follows.

\begin{code}
 tagShape :: (w -> w') -> Shape w -> Shape w'
 tagShape f s = s { shapeTail = f (shapeTail s) }
\end{code}
Embedding (or tagging) clearly does not disturb the reusable
definitions of functions on @Shape w@. However, the loop over
@scribble@ refers to the @draw@ operation, which is defined for
@RectangleDelta@ and @CircleDelta@, but not for the union over these two
types. We have to provide a trivial boilerplate for generalising @draw@:

\begin{code}
 instance (Draw a, Draw b) => Draw (Either a b)
  where
   draw = eitherShape draw draw
\end{code}
(This instance actually suffices for arbitrarily nested unions of @Shape@ subtypes.)
Here, @eitherShape@ is variation on the normal fold operation for
unions, i.e., @either@. We discriminate on the @Left@ vs.\ @Right@ cases
for the tail of a shape datum. This boilerplate operation is
independent of @draw@, but specific to @Shape@.

\begin{code}
 eitherShape :: (Shape w -> t) -> (Shape w' -> t) -> Shape (Either w w') -> t
 eitherShape f g s
   = case shapeTail s of
       (Left s')  -> f (s { shapeTail = s' })
       (Right s') -> g (s { shapeTail = s' })
\end{code}
The @Draw@ instance for @Either@ makes it clear that we use the union
type as an intersection type. We may only invoke a method on the union
only if we may invoke the method on either branch of the union. The
instance constraints make that fact obvious.

\paragraph{Assessment of the encoding}

\mbox{}

\begin{itemize}

\item
Again, the encoding ignores the encapsulation premise of the OO
paradigm: methods are not encapsulated along with the data.

\smallskip

\item
The encoding does not have the basic extensibility problem of the
previous section. We can introduce new kinds of shapes
without rewriting and recompiling \emph{type-specific} code.

\smallskip

\item
Some patterns of subtype-polymorphic code may require revision, though. For
instance all program points that insert into a subtype-polymorphic
collection or that downcast must agree on the formation of the union
type over specific subtypes. If a new subtype must be covered, then
the scattered applications of embedding operations must be revised.

\smallskip

\item 
We fail to put Haskell's type inference to work as far as object types
are concerned. We end up defining explicit datatypes for all
encoded classes. This is acceptable from a mainstream OO point of view
since nominal types (i.e., explicit types) dominate the OO
paradigm. However, in Haskell, we would like to do better by allowing
for inference of structural class and interface types. All subsequent
encodings of this section will share this problem. (By contrast,
\OOHaskell\ provides full structural type inference.)

\smallskip

\item
It is annoying enough that the formation of a subtype-polymorphic
collection requires explicit tagging of all elements; cf.\ @Left@ and
@Right@. What is worse, the tagging is done on the delta position of
@Shape@. This makes the scheme non-compositional: Each new base class
requires its own functions like @tagShape@ and @eitherShape@.

\smallskip

\item
The encoding of final and virtual methods differs essentially.  The
former are encoded as parametric polymorphic functions parameterised
in the extension type. Virtual methods are encoded as
type-class-bounded polymorphic functions overloaded in the extension
type. Changing a final method into virtual or vice versa triggers code
rewriting. This may be overcome by making all methods virtual (and
using default type class methods to reuse implementations). However,
this bias will increase the amount of boilerplate code such as the
instances for @Either@.

\smallskip

\item
The subtyping hierarchy leaks into the encoding of subtype-specific
accessors; cf.\ @setWidth@. The derivation chain from a base type
shows up as nesting depth in the record access pattern. One may factor
out these code patterns into access helpers and overload
them so that all accessors can be coded in a uniform way. This will
complicate the encoding, though.

\smallskip

\item
The approach is restricted to single inheritance.

\end{itemize}


\subsection{Functional objects, again with tail polymorphism}
\label{S:funcobj}

So far we have defined all methods as separate functions that process
``data records''. Hence, we ignored the OO encapsulation premise:
our data and methods were divorced from each other. Thereby,
we were able to circumvent problems of self references that tend to
occur in object encodings. Also, we avoided the classic dichotomy
`mutable vs.\ functional objects'. We will complement the picture by
exploring a functional object encoding (this section) and a mutable
object encoding (next section). We continue to use tail-polymorphic
records.

In a functional object encoding, object types are necessarily
recursive because all mutating methods are modelled as record
components that return ``self''. In fact, the type-theoretic technique
is to use equi-recursive types~\cite{PT94}.  We must use iso-recursive
types instead since Haskell lacks equi-recursive types.

Extensible shapes are modelled through the following recursive datatype:

\begin{code}
 data Shape w =
      Shape { getX      :: Int
            , getY      :: Int
            , setX      :: Int -> Shape w
            , setY      :: Int -> Shape w
            , moveTo    :: Int -> Int -> Shape w
            , rMoveTo   :: Int -> Int -> Shape w
            , draw      :: IO ()
            , shapeTail :: w
           }
\end{code}
This type reflects the complete interface of shapes, including
getters, setters, and more complex methods. The object constructor is
likewise recursive.  Recall that recursion models functional mutation,
i.e., the construction of a changed object:

\begin{code}
 shape x y d t
   = Shape { getX      = x
           , getY      = y
           , setX      = \x' -> shape x' y d t
           , setY      = \y' -> shape x y' d t
           , moveTo    = \x' y' -> shape x' y' d t
           , rMoveTo   = \deltax deltay -> shape (x+deltax) (y+deltay) d t
           , draw      = d x y
           , shapeTail = t
           }
\end{code}

As before, subtypes are modelled as instantiations of the base-type
record. That is, the @Rectangle@ record type is an instance of the
@Shape@ record type, where instantiation fixes the type @shapeTail@
somewhat:

\begin{code}
 type Rectangle w = Shape (RectangleDelta w)
\end{code}

\begin{code}
 data RectangleDelta w =
      RectangleDelta { getWidth'     :: Int 
                     , getHeight'    :: Int
                     , setWidth'     :: Int -> Rectangle w
                     , setHeight'    :: Int -> Rectangle w
                     , rectangleTail :: w
                     }
\end{code}
We used primed labels because we wanted to save the unprimed names for
the actual programmer API. The following implementations of the
unprimed functions hide the fact that rectangle records are
nested.

\begin{code}
 getWidth  = getWidth'  . shapeTail
 getHeight = getHeight' . shapeTail
 setWidth  = setWidth'  . shapeTail
 setHeight = setHeight' . shapeTail
\end{code}

The constructor for rectangles elaborates the constructor for shapes as follows:

\begin{code}
 rectangle x y w h
   = shape x y drawRectangle shapeTail
  where
   drawRectangle x y
     =  putStrLn ("Drawing a Rectangle at:("
     ++ (show x) ++ ","
     ++ (show y) ++ "), width "
     ++ (show w) ++ ", height "
     ++ (show h) )
   shapeTail 
     = RectangleDelta { getWidth'     = w 
                      , getHeight'    = h
                      , setWidth'     = \w' -> rectangle x y w' h
                      , setHeight'    = \h' -> rectangle x y w h'
                      , rectangleTail = ()
                      }
\end{code}
The encoding of the subclass @Circle@ can be derived likewise. (Omitted.)

This time, the scribble loop is set up as follows:

\begin{code}
 main =
       do
          let scribble = [ narrowToShape (rectangle 10 20 5 6)
                         , narrowToShape (circle 15 25 8)
                         ]
          mapM_ ( \x -> 
                    do
                       draw x
                       draw (rMoveTo x 100 100) )
                scribble
\end{code}

The interesting aspect of this encoding concerns the construction of
the @scribble@ list. We \emph{cast} or narrow the shapes of
different kinds to a common type. This is a general option, which we
could have explored in the previous section (where
we used embedding into a union type instead). Narrowing takes a shape
with an arbitrary tail and returns a shape with tail @()@:

\begin{code}
 narrowToShape :: Shape w -> Shape ()
 narrowToShape s = s { setX      = narrowToShape . setX s
                     , setY      = narrowToShape . setY s 
                     , moveTo    = \z -> narrowToShape . moveTo s z 
                     , rMoveTo   = \z -> narrowToShape . rMoveTo s z
                     , shapeTail = ()
                     }
\end{code}

\paragraph{Assessment of the encoding}

\mbox{}

\begin{itemize}

\item
The encoding is faithful to the encapsulation premise of the OO paradigm.

\smallskip

\item
The specific extensibility problem of the `union type' approach is
resolved (cf.\ assessment Sec.~\ref{S:burton}). Code that accesses a
subtype-polymorphic collection does not need to be revised when new
subtypes are added elsewhere in the program. The `narrowing'
approach frees the programmer from commitment to a specific union type.

\smallskip

\item 
The narrowing approach (unlike the union-type one) does not permit
downcasting.

\smallskip

\item
The implementation of the narrowing operation is base-type-specific
just as the earlier embedding helpers for union types. Boilerplate code
of that kind is, of course, not required from programmers in mainstream
OO languages.

\end{itemize}


\subsection{Mutable objects, again with tail polymorphism}
\label{S:mutable}

We also review an object encoding for mutable objects, where we employ
@IORef@s of the @IO@ monad to enable object state~---~as is the case
for \OOHaskell. The functions (``methods'') in a record
manipulate the state through @IORef@ operations.  We continue to use
tail-polymorphic records.

Extensible shapes are modelled through the following type:

\begin{code}
 data Shape w =
      Shape { getX      :: IO Int
            , getY      :: IO Int
            , setX      :: Int -> IO ()
            , setY      :: Int -> IO ()
            , moveTo    :: Int -> Int -> IO ()
            , rMoveTo   :: Int -> Int -> IO ()
            , draw      :: IO ()
            , shapeTail :: w
            }
\end{code}

The result type of all methods is wrapped in the @IO@ monad so that
all methods may have side effects, if necessary.  One may wonder
whether this is really necessary for getters.  Even for a getter, we
may want to add memoisation or logging, when we override the method in
a subclass, in which case a non-monadic result type would be too
restrictive.

The object generator (or constructor) for shapes is parameterised in
the initial shape position @x@ and @y@, in a concrete implementation
of the abstract method @draw@, in the @tail@ of the record to be
contributed by the subtype, and in @self@ so to enable open
recursion. The latter lets subtypes override method defined in
@shape@. (We will illustrate overriding shortly.)

\begin{code}
 shape x y concreteDraw tail self
   = do
        xRef  <- newIORef x
        yRef  <- newIORef y
        tail' <- tail
        returnIO Shape
                  { getX      = readIORef xRef
                  , getY      = readIORef yRef
                  , setX      = \x' -> writeIORef xRef x'
                  , setY      = \y' -> writeIORef yRef y'
                  , moveTo    = \x' y' -> do { setX self x'; setY self y' }
                  , rMoveTo   = \deltax deltay -> 
                                  do x <- getX self
                                     y <- getY self
                                     moveTo self (x+deltax) (y+deltay)
                  , draw      = concreteDraw self
                  , shapeTail = tail' self
                  }
\end{code}

The type declarations for rectangles are the following:

\begin{code}
 type Rectangle w = Shape (RectangleDelta w)
\end{code}

\begin{code}
 data RectangleDelta w =
      RectangleDelta { getWidth'     :: IO Int 
                     , getHeight'    :: IO Int
                     , setWidth'     :: Int -> IO ()
                     , setHeight'    :: Int -> IO ()
                     , rectangleTail :: w
                     }
\end{code}
Again, we define unprimed names to hide the nested status of the
rectangle API:

\begin{code}
 getWidth  = getWidth'  . shapeTail
 getHeight = getHeight' . shapeTail
 setWidth  = setWidth'  . shapeTail
 setHeight = setHeight' . shapeTail
\end{code}

We reveal the object generator for rectangles step by step.

\begin{code}
 rectangle x y w h
   = shape x y drawRectangle shapeTail
  where
    -- to be cont'd
\end{code}
We invoke the generator @shape@, passing on the normal constructor
arguments @x@ and @y@, a rectangle-specific draw method, and the tail
for the rectangle API. We do \emph{not} yet fix the @self@ reference,
thereby allowing for further subtyping of @rectangle@. We define the
@draw@ method as follows, resorting to C++-like syntax, @<<@, for daisy
chaining output:

\begin{code}
   drawRectangle self =  
        putStr "Drawing a Rectangle at:(" <<
        getX self << ls "," << getY self <<
        ls "), width " << getWidth self <<
        ls ", height " << getHeight self <<
        ls "\n"
\end{code}
Finally, the following is the rectangle part of the shape object:

\begin{code}
   shapeTail
     = do 
          wRef <- newIORef w
          hRef <- newIORef h
          returnIO ( \self -> 
             RectangleDelta
                 { getWidth'     = readIORef wRef 
                 , getHeight'    = readIORef hRef
                 , setWidth'     = \w' -> writeIORef wRef w'
                 , setHeight'    = \h' -> writeIORef hRef h'
                 , rectangleTail = ()
                 } )
\end{code}
The overall subtype derivation scheme is at ease with
overriding methods in subtypes. We illustrate this capability by
temporarily assuming that the @draw@ method is not abstract. So we may
revise the constructor for shapes as follows:

\begin{code}
 shape x y tail self
   = do
        xRef  <- newIORef x
        yRef  <- newIORef y
        tail' <- tail
        returnIO Shape
                  { -- ... as before but we deviate for draw ...
                  , draw = putStrLn "Nothing to draw"
                  }
\end{code}
We override @draw@ when constructing rectangles:

\begin{code}
 rectangle x y w h self
   = do 
        super <- shape x y shapeTail self
        returnIO super { draw = drawRectangle self }
\end{code}

As in the previous section, we use @narrowToShape@ when building a
list of different shapes. Actual object construction ties the recursive
knot for the self references with @mfix@. Hence, @mfix@ is our
operator ``new''.

\begin{code}
main =
      do
         s1 <- mfix $ rectangle 10 20 5 6
         s2 <- mfix $ circle 15 25 8
         let scribble = [ narrowToShape s1
                        , narrowToShape s2
                        ]
         mapM_ ( \x -> 
                   do
                      draw x
                      rMoveTo x 100 100
                      draw x )
               scribble
\end{code}
The narrow operation is trivial this time:

\begin{code}
 narrowToShape :: Shape w -> Shape ()
 narrowToShape s = s { shapeTail = () } 
\end{code}

We just ``chop off'' the tail of shape objects. We may no longer use
any rectangle- or circle-specific methods. One may say that chopping
off the tail makes the fields in the tail and the corresponding
methods \emph{private}. The openly recursive
methods, in particular @draw@, had access to self that characterised
the whole object, \emph{before the chop off}. The narrow operation
becomes (potentially much) more involved or infeasible once we
consider self-returning methods, binary methods, co- and
contra-variance, and other advanced OO idioms.

\paragraph{Assessment of the encoding}

This encoding is actually very close to \OOHaskell\ except that the
former uses explicitly declared, non-extensible record types.  As a
result, the encoding requires substantial boilerplate code (to account
for type extension) and subtyping is explicit. Furthermore,
\OOHaskell\ leverages type-level programming to lift restrictions
like the limited narrowing capabilities.


\subsection{Subtypes as composed record types with overloading}
\label{S:objcomp}

Many problems of tail-polymorphic record types prompt us to consider
an alternative. We now compose record types for subtypes and use type
classes to represent the actual subtype relationships.  Such use of
type classes has first been presented in~\cite{SPJ01} for encoding of
OO interface polymorphism in Haskell. We generalise this technique for
class inheritance.

The compositional approach can be described as follows:
\begin{itemize}

\item The data part of an OO class amounts to a record type.
\item Each such record type includes components for superclass data
\item The interface for each OO class amounts to a Haskell type class.
\item OO superclasses are mapped to Haskell superclass constraints.
\item Reusable OO method implementations are mapped to default methods.
\item A subtype is implemented as a type-class instance.

\end{itemize}

We begin with the record type for the data part of the
@Shape@ class:

\begin{code}
 data ShapeData =
      ShapeData { valX :: Int
                , valY :: Int }
\end{code}
For convenience, we also provide a constructor:

\begin{code}
 shape x y = ShapeData { valX = x
                       , valY = y }
\end{code}

We define a type class @Shape@ that models the OO interface for shapes:

\begin{code}
 class Shape s
  where
   getX       :: s -> Int
   setX       :: Int -> s -> s
   getY       :: s -> Int
   setY       :: Int -> s -> s
   moveTo     :: Int -> Int -> s -> s
   rMoveTo    :: Int -> Int -> s -> s
   draw       :: s -> IO ()
   -- to be cont'd
\end{code}

We would like to provide reusable definitions for most of these
methods (except for @draw@ of course). In fact, we would like to
define the accessors to shape data once and for all. To this end, we
need additional helper methods. While it is clear how to
define accessors on @ShapeData@, we must provide generic definitions
that are able to handle records that \emph{include} @ShapeData@ as one
of their (immediate or non-immediate) components.  This leads to the
following two helpers:

\begin{code}
 class Shape s
  where
   -- cont'd from earlier
   readShape  :: (ShapeData -> t)         -> s -> t
   writeShape :: (ShapeData -> ShapeData) -> s -> s
\end{code}
which let us define generic shape accessors:

\begin{code}
 class Shape s
  where
   -- cont'd from earlier
   getX       =  readShape valX
   setX i     =  writeShape (\s -> s  { valX = i })
   getY       =  readShape valY
   setY i     =  writeShape (\s -> s  { valY = i })
   moveTo x y =  setY y . setX x 
   rMoveTo deltax deltay s = moveTo x y s
    where
      x = getX s + deltax
      y = getY s + deltay
\end{code}

We do \emph{not} define an instance of the @Shape@ class for
@ShapeData@ because the original shape class was abstract
due to the purely virtual @draw@ method. As we move to rectangles,
we define their data part as follows:

\begin{code}
 data RectangleData =
      RectangleData { valShape  :: ShapeData
                    , valWidth  :: Int 
                    , valHeight :: Int
                    }
\end{code}
The rectangle constructor also invokes the shape constructor:

\begin{code}
 rectangle x y w h
  = RectangleData { valShape  = shape x y
                  , valWidth  = w
                  , valHeight = h
                  }
\end{code}

``A rectangle is a shape.'' We provide access to the shape part as follows:

\begin{code}
 instance Shape RectangleData
  where
   readShape f    = f . valShape
   writeShape f s = s { valShape = readShape f s } 
   -- to be cont'd
\end{code}
We also implement the @draw@ method.

\begin{code}
   -- instance Shape RectangleData cont'd
   draw s
     =   putStrLn ("Drawing a Rectangle at:("
     ++ (show (getX s)) ++ ","
     ++ (show (getY s)) ++ "), width "
     ++ (show (getWidth s)) ++ ", height "
     ++ (show (getHeight s)))
\end{code}

We also need to define a Haskell type class for the OO class of
rectangles. OO subclassing coincides in Haskell type-class subclassing.

\begin{code}
 class Shape s => Rectangle s
  where
   -- to be cont'd
\end{code}
The type class is derived from the corresponding OO class just as we
explained for the base class of shapes. The class defines the `normal'
interface of rectangles and access helpers.

\begin{code}
   -- class Rectangle cont'd
   getWidth       :: s -> Int
   getWidth       =  readRectangle valWidth
   setWidth       :: Int -> s -> s
   setWidth i     =  writeRectangle (\s -> s  { valWidth = i })
   getHeight      :: s -> Int
   getHeight      =  readRectangle valHeight
   setHeight      :: Int -> s -> s    
   setHeight i    =  writeRectangle (\s -> s  { valHeight = i })
\end{code}

\begin{code}
   readRectangle  :: (RectangleData -> t)         -> s -> t
   writeRectangle :: (RectangleData -> RectangleData) -> s -> s
\end{code}

``A rectangle is (nothing but) a rectangle.''

\begin{code}
 instance Rectangle RectangleData
  where
   readRectangle  = id
   writeRectangle = id
\end{code}

The subclass for circles can be encoded in the same way.

The scribble loop can be performed on tagged rectangles and circles:

\begin{code}
 main =
       do
          let scribble = [ Left  (rectangle 10 20 5 6)
                         , Right (circle 15 25 8)
                         ]
          mapM_ ( \x -> 
                    do
                       draw x
                       draw (rMoveTo 100 100 x)
                )
                scribble
\end{code}
We attach @Left@ and @Right@ tags at the top-level this time. Such
simple tagging was not possible with the tail-polymorphic encodings.  We
still need an instance for @Shape@ that covers tagged shapes:

\begin{code}
 instance (Shape a, Shape b) => Shape (Either a b)
  where
   readShape  f = either (readShape f)  (readShape f)
   writeShape f = bimap  (writeShape f) (writeShape f)
   draw         = either draw draw
\end{code}

The bi-functorial map, @bimap@, pushes @writeShape@ into the tagged
values. The @Either@-specific fold operation, @either@, pushes
@readShape@ and @draw@ into the tagged values. For completeness, we
recall the relevant facts about bi-functors and folds on @Either@:

\begin{code}
 class BiFunctor f where
   bimap :: (a -> b) -> (c -> d) -> f a c -> f b d
\end{code}

\begin{code}
 instance BiFunctor Either where
   bimap f g (Left  x)  = Left (f x)
   bimap f g (Right x') = Right (g x')
\end{code}

\begin{code}
 either :: (a -> c) -> (b -> c) -> Either a b -> c
 either f g (Left x)  =  f x
 either f g (Right y) =  g y
\end{code}

We should mention a minor but useful variation, which avoids the
explicit attachment of tags when \emph{inserting} into a
subtype-polymorphic collection.  We use a special cons operation,
@consEither@, which replaces the normal list constructor @(:)@:

\begin{code}
          -- ... so far ...
          let scribble = [ Left  (rectangle 10 20 5 6)
                         , Right (circle 15 25 8)
                         ]
\end{code}

\begin{code}
          -- ... liberalised notation ...
          let scribble = consEither
                        (rectangle 10 20 5 6)
                        [circle 15 25 8]
\end{code}

\begin{code}
 -- A union-constructing cons operation
 consEither :: h -> [t] -> [Either h t]
 consEither h t@(_:_) = Left h : map Right t 
 consEither _ _ = error "Cannot cons with empty tail!"
\end{code}

\paragraph{Assessment of the encoding}

\mbox{}

\begin{itemize}

\item 
This approach is highly systematic and general. For instance, multiple
inheritance is immediately possible. One may argue that this approach
does not directly encode OO class inheritance. Rather, it mimics
\emph{object composition}. One might indeed convert native OO programs, prior
to encoding, so that they do not use class inheritance, but they use
interface polymorphism combined with (manually coded) object
composition instead.

\smallskip

\item 
A fair amount of boilerplate code is required (cf.\ @readShape@ and
@writeShape@).  Also, each \emph{transitive} subtype relationship
requires surprising boilerplate. For example, let us assume
an OO class @FooBar@ that is a subclass of
@Rectangle@. The transcription to Haskell would involve three
instances: one for the type class that is dedicated to @FooBar@
(``Ok''), one for @Rectangle@ (still ``Ok'' except the scattering 
of implementation), and one for @Shape@ (``annoying'').

\smallskip

\item
The union-type technique improved compared to Sec.~\ref{S:burton}.
The top-level tagging scheme eliminates the need for tagging helpers
that are specific to the object types.  Also, the @consEither@
operation relieves us from the chore of explicitly writing sequences
of tags.  However, we must assume that we insert into a non-empty
list, and we also must accept that the union type increases for each
new element in the list~---~no matter how many different element types
are encountered. Also, if we want to downcast from the union type, we
still need to know its exact layout.  To lift these restrictions, we
have to engage into proper type-class-based programming.

\end{itemize}


\subsection{Variation~---~existential quantification}
\label{S:exists}

So far we have restricted ourselves to Haskell~98. We now turn to
common extensions of Haskell~98, in an attempt to improve on the
problems that we have encountered. In upshot, we cannot spot obvious
ways for improvement.

Our first attempt is to leverage existential quantification for the
implementation of subtype-polymorphic
collections. Compared to the earlier @Either@-based approach, we
homogenise shapes by making them opaque~\cite{CW85} as opposed to
embedding them into the union type.  This use of existentials could be
combined with various object encodings; we illustrate it here for the
specific encoding from the previous section.

We define an existential envelope for shape data.

\begin{Verbatim}[fontsize=\small,commandchars=\\\{\}]
 data OpaqueShape = forall x. Shape x => HideShape x
\end{Verbatim}

``Opaque shapes are (still) shapes.'' Hence, a @Shape@ instance:

\begin{code}
 instance Shape OpaqueShape 
  where
   readShape  f (HideShape x) = readShape  f x
   writeShape f (HideShape x) = HideShape $ writeShape f x
   draw         (HideShape x) = draw x
\end{code}

%
When building the scribble list, we place shapes in the envelope.

\begin{code}
 let scribble = [ HideShape (rectangle 10 20 5 6)
                , HideShape (circle 15 25 8)
                ]
\end{code}

\paragraph{Assessment of the encoding}

\mbox{}

\begin{itemize}

\item 
Compared to the `union type' approach, programmers do not have to
invent union types each time they need to homogenise different
subtypes. Instead, \emph{all} shapes are tagged by @HideShape@. The
`narrowing' approach was quite similar, but it required boilerplate.

\smallskip

\item
We face a new problem. Existential quantification limits type
inference.

\end{itemize}
We see that in the definition of @OpaqueShape@; viz.\ the explicit
constraint @Shape@.  It is mandatory to constraint the quantifier by
\emph{all} subtypes whose methods may be invoked. A reader may notice
the similar problem for the `union type' approach, which required
@Shape@ constraints in the instance

\begin{code}
 instance (Shape a, Shape b) => Shape (Either a b) where ...	
\end{code}
That instance, however, was merely a convenience. We could have 
disposed of it and used the fold operation @either@ explicitly in the
scribble loop:

\begin{code}
 main =
      do
         let scribble = [ Left  (rectangle 10 20 5 6)
                        , Right (circle 15 25 8)
                        ]
         mapM_ (either scribbleBody scribbleBody) scribble
 where
   scribbleBody x = do
                   draw x
                   draw (rMoveTo 100 100 x) 
\end{code}

By contrast, the explicit constraint for the existential envelope
cannot be eliminated.  Admittedly, the loss of type inference is a
nuance in this specific example.  In general, however, this weakness
of existentials is quite annoying. It is intellectually dissatisfying
since type inference is one of the added values of an (extended)
Hindley/Milner type system, when compared to mainstream OO
languages. Worse than that, the kind of constraints in the example are
\emph{not} necessary in mainstream OO languages (without type inference),
because these constraints deal with subtyping, which is normally
\emph{implicit}.

We do not use existentials in \OOHaskell.


\subsection{Variation~---~heterogeneous collections}
\label{S:hetero}

{\sloppypar 

We continue with our exploration of common extensions of Haskell~98.
In fact, we will offer another option for the difficult problem of
subtype-polymorphic collections. We recall that all previously
discussed techniques aimed at making it possible to construct a normal
homogeneous Haskell list in the end. This time, we will engage into the
construction of a \emph{heterogeneous} collection in the first
place. To this end, we leverage techniques as described by us in the
\HList\ paper~\cite{HLIST-HW04}. Heterogeneous collections rely on
multi-parameter classes~\cite{CHO92,MPJ92,MPJ95,PJJM97} with
functional dependencies~\cite{MPJ00,DPJSS04}.

}

Heterogeneous lists are constructed with dedicated constructors
@HCons@ and @HNil@~---~analogues of @(:)@ and @[]@. One may
think of a heterogeneous list type as a nested binary product, where
@HCons@ corresponds to @(,)@ and @HNil@ to @()@. We use special
\HList\ functions to process the heterogeneous lists; the
example requires a map operation. The scribble loop is now encoded as
follows:

\begin{code}
main =
      do
         let scribble =  HCons (rectangle 10 20 5 6)
                        (HCons (circle 15 25 8)
                         HNil)
         hMapM_ (undefined::ScribbleBody) scribble
\end{code}
The operation @hMapM_@ is the heterogeneous variation on the normal
monadic map @mapM_@. The function argument for the map cannot be given
inline; instead we pass a proxy @undefined::ScribbleBody@. This
detour is necessary due to technical reasons that are related to the
combination of rank-2 polymorphism and type-class-bounded
polymorphism.\footnote{\small A heterogeneous map function can
encounter entities of different types. Hence, its argument function
must be polymorphic on its own (which is different from the normal map
function for lists). The argument function typically uses
type-class-bounded polymorphic functions to process the entities of
different types. The trouble is that the map function cannot possibly
anticipate all the constraints required by the different uses of the
map function. The type-code technique moves the constraints from the
type of the heterogeneous map function to the interpretation site of
the type codes.}

The type code for the body of the scribble loop is defined by a
trivial datatype:

\begin{code}
 data ScribbleBody -- No constructors needed; non-Haskell 98
\end{code}

The heterogeneous map function is constrained by the @Apply@ class,
which models interpretation of function codes like @ScribbleBody@.
Here is the @Apply@ class and the instance dedicated to
@ScribbleBody@:

\begin{code}
 class  Apply f a r | f a -> r
  where apply :: f -> a -> r
\end{code}

\begin{code}
 instance Shape s => Apply ScribbleBody s (IO ())
  where
   apply _ x = 
      do
         draw x
         draw (rMoveTo 100 100 x)
\end{code}

\paragraph{Assessment of the encoding}

\mbox{}

\begin{itemize}

\item 
This approach eliminates all effort for inserting elements into a
collection.

\smallskip

\item 
The approach comes with heavy surface encoding; cf.\ type code
@ScribbleBody@.

\smallskip

\item
This encoding is at odds with type inference~---~just as in the case of
existentials. That is, the @Apply@ instance must be explicitly
constrained by the interfaces that are going to be relied upon in the
body of the scribble loop. Again, the amount of explicit typing is not
yet disturbing in the example at hand, but it is an intrinsic weakness
of the encoding. The sort of required explicit typing goes beyond
standard OO programming practise.

\end{itemize}


\section{Type-agnostic OOHaskell idioms}
\label{S:OOHaskell1}

We will now systematically develop all important \OOHaskell\
programming idioms. In this section, we will restrict ourselves to
`type-agnostic' idioms, as to clearly substantiate that most
\OOHaskell\ programming does not require type declarations, type
annotations, explicit casts for object types~---~thanks to Haskell's
type inference and its strong support for polymorphism. The remaining,
`type-perceptive' \OOHaskell\ idioms (including a few advanced topics
related to subtyping) are described in the subsequent section.

In both sections, we adopt the following style. We illustrate the
OO idioms and describe the technicalities of encoding.  We
highlight strengths of \OOHaskell: support for the traditional OO
idioms as well as extra features due to the underlying record
calculus, and first-class status of labels, methods and
classes. Finally, we illustrate the overall programmability of a typed
OO language design in Haskell.

As a matter of style, we somewhat align the presentation of \OOHaskell\ 
with the OCaml object tutorial. Among the many OO systems that are
based on open records (Perl, Python, Javascript, Lua, etc.), OCaml
stands out because it is statically typed (just as \OOHaskell). Also,
OCaml (to be precise, its predecessor {ML-ART}) is close to \OOHaskell\ 
in terms of motivation: both aim at the introduction of objects as a
library in a strongly-typed functional language with type
inference. The implementation of the libraries and the sets of
features used or required are quite different (cf.\ Sec.~\ref{S:MLART}
for a related work discussion), which makes a comparison even more
interesting. Hence, we draw examples from the OCaml object tutorial,
to specifically contrast OCaml and \OOHaskell\ code and to demonstrate
the fact that OCaml examples are expressible in \OOHaskell, roughly in
the same syntax, based on direct, local translation. We also use the
OCaml object tutorial because it is clear, comprehensive and concise.


\subsection{Objects as records}

Quoting from~\cite{OCaml}[\S\,3.2]:\footnote{While quoting portions of
the OCaml tutorial, we take the liberty to rename some identifiers and
to massage some subminor details.}

\begin{quote}\itshape\small
``The class @point@ below defines one instance variable @varX@ and two
methods @getX@ and @moveX@. The initial value of the instance variable
is @0@. The variable @varX@ is declared mutable. Hence, the method
@moveX@ can change its value.''
\end{quote}

\begin{code}
 class point =
   object
     val mutable varX = 0
     method getX      = varX
     method moveX d   = varX <- varX + d
   end;;
\end{code}


\subsubsection{First-class labels}

The transcription to \OOHaskell\ starts with the declaration of all the
labels that occur in the OCaml code. The \HList\ library readily
offers 4 different models of labels. In all cases, labels are Haskell
values that are distinguished by their Haskell \emph{type}. We choose
the following model:
\begin{itemize}
\item
The value of a label is ``\undefined''.
\item
The type of a label is a \emph{proxy} for an \emph{empty} type (empty
except for ``\undefined'').\footnote{It is a specific GHC extension of
Haskell~98 to allow for datatypes without any constructor
declarations. Clearly, this is a minor issue because one could always
declare a dummy constructor that is not used by the program.}
\end{itemize}

\begin{code}
 data VarX;  varX  = proxy :: Proxy VarX 
 data GetX;  getX  = proxy :: Proxy GetX
 data MoveX; moveX = proxy :: Proxy MoveX
\end{code}
where proxies are defined as

\begin{Verbatim}[fontsize=\small,commandchars=\\\{\}]
 data Proxy e      -- \cmt{A proxy type is an empty phantom type.}
 proxy :: Proxy e  -- \cmt{A proxy value is just ``\undefined''.}
 proxy = \undefined
\end{Verbatim}
Simple syntactic sugar can significantly reduce the length of the
one-liners for label declaration should this become an issue. For
instance, we may think of the above lines just as follows:

\begin{code}
 -- Syntax extension assumed; label is a new keyword.
 label varX
 label getX
 label moveX
\end{code}

The \emph{explicit} declaration of \OOHaskell\ labels blends well with
Haskell's scoping rules and its module concept. Labels can be private
to a module, or they can be exported, imported, and shared. All models of
\HList\ labels support labels as first-class citizens. In particular,
we can pass them to functions. The ``labels as type proxies'' idea is
the basis for defining record operations since we can thereby
\emph{dispatch} on labels in type-level functionality. We
will get back to the record operations shortly.


\subsubsection{Mutable variables}

The OCaml @point@ class is transcribed to \OOHaskell\ as follows:

\begin{code}
 point = 
   do
      x <- newIORef 0
      returnIO
        $  varX  .=. x
       .*. getX  .=. readIORef x
       .*. moveX .=. (\d -> do modifyIORef x (+d))
       .*. emptyRecord
\end{code}

{\sloppypar

The \OOHaskell\ code clearly mimics the OCaml code. While we use
Haskell's @IORef@s to model mutable variables, we do not use any magic
of the IO monad. We could as well use the simpler ST monad, which is
very well formalised~\cite{LPJ95}. The source distribution for the
paper illustrates the ST option.

}

The Haskell representation of the @point@ class stands revealed as a
value binding declaration of a monadic record type. The @do@ sequence
first creates an @IORef@ for the mutable variable, and then returns a
record for the new @point@ object. In general, the \OOHaskell\ records
provide access to the public methods of an object and to the @IORef@s
for \emph{public} mutable variables. We will often call all record
components of \OOHaskell's objects just `methods'. In the example,
@varX@ is public, just as in the original OCaml code. In \OOHaskell, a
\emph{private} variable would be encoded as an @IORef@ that is not made
available through a record component. (Private variables were explored
in the shapes example.)


\subsubsection{HList records}

We may ask Haskell to tell us the inferred type of @point@:

\begin{code}
 ghci> :t point
 point :: IO (Record (HCons (Proxy MutableX, IORef Integer)
                     (HCons (Proxy GetX, IO Integer)
                     (HCons (Proxy Move, Integer -> IO ())
                      HNil))))
\end{code}
The type reveals the use of \HList's extensible
records~\cite{HLIST-HW04}. We explain some details about
\HList, as to make the present paper self-contained. The inferred
type shows that records are represented as \emph{heterogeneous
label-value pairs}, which are \emph{promoted to a proper record type}
through the type-constructor @Record@.

\begin{code}
 -- HList constructors
 data HNil      = HNil      -- empty heterogeneous list
 data HCons e l = HCons e l -- non-empty heterogeneous list
\end{code}

\begin{code}
 -- Sugar for forming label-value pairs
 infixr 4 .=.
 l .=. v = (l,v)
\end{code}

\begin{code}
 -- Record type constructor
 newtype Record r = Record r
\end{code}

The constructor @Record@ is opaque for the library user. Instead, the
library user (and most of the library code itself) relies upon a
constrained constructor:

\begin{code}
 -- Record value constructor
 mkRecord :: HRLabelSet r => r -> Record r
 mkRecord = Record
\end{code}
The constraint @HRLabelSet r@ statically assures that all labels are
pairwise distinct as this is a necessary precondition for a list of
label-value pairs to qualify as a record. (We omit the routine
specification of @HRLabelSet r@~\cite{HLIST-HW04}.) We can now
implement @emptyRecord@, which was used in the definition of @point@:

\begin{code}
 emptyRecord = mkRecord HNil
\end{code}

The record extension operator, @(.*.)@, is a constrained variation on
the heterogeneous cons operation, @HCons@: we need to make sure that
the newly added label-value pair does not violate the uniqueness
property for the labels. This is readily expressed by wrapping the
unconstrained cons term in the constrained record constructor:

\begin{code}
 infixr 2 .*.
 (l,v) .*. (Record r) = mkRecord (HCons (l,v) r)
\end{code}


\subsubsection{OO test cases}

We want to instantiate the @point@ class and invoke some methods.  We
begin with \emph{an OCaml session}, which shows some inputs and the
responses from the OCaml interpreter:\footnote{OCaml's prompt is
indicated by a leading character ``\#''.\\ Method invocation is
modelled by the infix operator ``\#''.\\ The lines with leading
``\texttt{val}'' or ``--'' are the responses from the interpreter.}

\begin{code}
 # let p = new point;;
 val p : point = <obj>
 # p#getX;;
 - : int = 0
 # p#moveX 3;;
 - : unit = ()
 # p#getX;;
 - : int = 3
\end{code}

In Haskell, we capture this program in a monadic @do@ sequence because
method invocations can involve IO effects including the mutation of
objects. We denote method invocation by @(#)@, just as in OCaml; this
operation is a plain record look-up. Hence:

\begin{code}
 myFirstOOP =
  do
     p <- point -- no need for new!
     p # getX >>= Prelude.print
     p # moveX $ 3
     p # getX >>= Prelude.print
\end{code}

\OOHaskell\ and OCaml agree:

\begin{code}
 ghci> myFirstOOP
 0
 3
\end{code}

For completeness we outline the definition of ``@#@'':

\begin{code}
 -- Sugar operator
 infixr 9 #
 obj # feature = hLookupByLabel feature obj
\end{code}

\begin{code}
 -- Type-level operation for look-up
 class HasField l r v | l r -> v
  where
   hLookupByLabel:: l -> r -> v
\end{code}
This operation performs type-level (and value-level) traversal of the
label-value pairs, looking up the value component for a given label
from the given record, while using the label type as a key. We recall
that the term `field' (cf.\ @HasField@) originates from record
terminology. In \OOHaskell, all `fields' are `methods'.  (We omit the
routine specification of @HasField@~@l@~@r@~@v@~\cite{HLIST-HW04}.) 
The class declaration reveals that \HList\ (and thereby \OOHaskell)
relies on multi-parameter classes~\cite{CHO92,MPJ92,MPJ95,PJJM97} with
functional dependencies~\cite{MPJ00,DPJSS04}.


\subsection{Object generators}

In class-based, mainstream OO languages, the construction of new class
instances is normally regulated by so-called constructor methods. In
OOHaskell, instances are created by a function that serves as an
object generator. The function can be seen as the embodiment of the
class itself. The @point@ computation defined above is a trivial
example of an object generator.


\subsubsection{Constructor arguments}

Quoting from~\cite{OCaml}[\S\,3.1]:

\begin{quote}\itshape\small
``The class @point@ can also be abstracted over the initial value of
@varX@.  The parameter @x_init@ is, of course, visible in the whole
body of the definition, including methods. For instance, the method
@getOffset@ in the class below returns the position of the object
relative to its initial position.''
\end{quote}

\begin{code}
 class para_point x_init =
   object
     val mutable varX = x_init
     method getX      = varX
     method getOffset = varX - x_init
     method moveX d   = varX <- varX + d
   end;;
\end{code}

In \OOHaskell, objects are created as the result of monadic
computations producing records. We can parameterise these computations
by turning them into functions, object generators, which take
construction parameters as arguments. For instance, the
parameter @x_init@ of the OCaml class @para_point@ ends up as a plain
function argument:

\begin{code}
 para_point x_init
   = do
        x <- newIORef x_init
        returnIO
          $  varX      .=. x
         .*. getX      .=. readIORef x
         .*. getOffset .=. queryIORef x (\v -> v - x_init)
         .*. moveX     .=. (\d -> modifyIORef x (+d))
         .*. emptyRecord
\end{code}


\subsubsection{Construction-time computations}

Quoting from~\cite{OCaml}[\S\,3.1]:

\begin{quote}\itshape\small
``Expressions can be evaluated and bound before defining the object
body of the class. This is useful to enforce invariants. For instance,
points can be automatically adjusted to the nearest point on a grid,
as follows:''
\end{quote}

\begin{code}
 class adjusted_point x_init =
   let origin = (x_init / 10) * 10 in
   object
     val mutable varX = origin
     method getX      = varX
     method getOffset = varX - origin
     method moveX d   = varX <- varX + d
   end;;
\end{code}

In \OOHaskell, we follow the suggestion from the OCaml tutorial:
we use local let bindings to carry out the constructor computations
``prior'' to returning the constructed object:

\begin{code}
 adjusted_point x_init
   = do
        let origin = (x_init `div` 10) * 10
        x <- newIORef origin
        returnIO
          $  varX      .=. x
         .*. getX      .=. readIORef x
         .*. getOffset .=. queryIORef x (\v -> v - origin)
         .*. moveX     .=. (\d -> modifyIORef x (+d))
         .*. emptyRecord
\end{code}
%
That ``prior'' is not meant in a temporal sense: \OOHaskell\ remains
a non-strict language, in contrast to OCaml.


\subsubsection{Implicitly polymorphic classes}

A powerful feature of \OOHaskell\ is implicit polymorphism for
classes. For instance, the class |para_point| is
polymorphic with regard to the point's coordinate~---~without our
contribution. This is a fine difference between the OCaml model and
our \OOHaskell\ transcription. In OCaml's definition of @para_point@, the
parameter @x_init@ was of the type @int@~---~because the operation
@(+)@ in OCaml can deal with integers only. The \OOHaskell\ points are
polymorphic~---~a point's coordinate can be any @Num@-ber, for
example, an @Int@ or a @Double@. Here is an example to illustrate
that:

\begin{code}
 myPolyOOP =
    do
       p  <- para_point (1::Int)
       p' <- para_point (1::Double)
       p  # moveX $ 2
       p' # moveX $ 2.5
       p  # getX >>= Prelude.print
       p' # getX >>= Prelude.print
\end{code}

The \OOHaskell\ points are actually \emph{bounded} polymorphic. The
point coordinate may be of any type that implements addition. Until
very recently, one could not express this in Java and in
C\#. Expressing bounded polymorphism in C++ is possible with
significant contortions. In (OO)Haskell, we did not have to do
anything at all. Bounded polymorphism (aka, generics) is available in
Ada95, Eiffel and a few other languages. However, in those languages,
the polymorphic type and the type bounds must be declared
\emph{explicitly}. (There are ongoing efforts to add some specific bits of type
inference to new versions of mainstream OO languages.) In (OO)Haskell,
the type system \emph{infers} the (bounded) polymorphism on its own,
in full generality.

The implicit polymorphism of OOHaskell does not injure static typing.
If we confuse @Int@s and @Double@s in the above code, e.g., by attempting
``@p@~@#@~@moveX@~@$@~@2.5@'',
then we get a type error saying that @Int@ is not the same as
@Double@. In contrast, the poor men's implementation of polymorphic
collections, e.g., in Java @<@ 1.5, which up-casts an element to the
most general @Object@ type when inserting it into the collection,
requires runtime-checked downcasts when accessing elements.


\subsubsection{Nested object generators}
\label{S:nested}

Quoting from~\cite{OCaml}[\S\,3.1]:

\begin{quote}\itshape\small
``The evaluation of the body of a class only takes place at object
creation time.  Therefore, in the following example, the instance
variable @varX@ is initialised to different values for two different
objects.''
\end{quote}

\begin{code}
 let x0 = ref 0;;
\end{code}

\begin{code}
 class incrementing_point :
   object
     val mutable varX = incr x0; !x0
     method getX      = varX
     method moveX d   = varX <- varX + d
   end;;
\end{code}
We test this new class at the OCaml prompt:

\begin{code}
 # new incrementing_point#getX;;
 - : int = 1
 # new incrementing_point#getX;;
 - : int = 2
\end{code}

The variable |x0| can be viewed as a ``class variable'', belonging to
a \emph{class object}. Recall that classes are represented
by object generators in \OOHaskell. Hence to build a class object we need 
a nested object generator:

\begin{code}
 incrementing_point = 
   do 
      x0 <- newIORef 0
      returnIO (
        do modifyIORef x0 (+1)
           x <- readIORef x0 >>= newIORef
           returnIO
             $  varX  .=. x
            .*. getX  .=. readIORef x
            .*. moveX .=. (\d -> modifyIORef x (+d))
            .*. emptyRecord)
\end{code}
%
We can nest generators to any depth since we just use normal Haskell
scopes. In the example, the outer level does the computation for the
point template (i.e., ``class''); the inner level constructs points
themselves. Here is a more suggestive name for the nested generator:

\begin{code}
 makeIncrementingPointClass = incrementing_point
\end{code}

This (trivial) example demonstrates that classes in \OOHaskell\ 
are really first-class citizens. We can pass classes as arguments
to functions and return them as results. In the following code fragment, we
create a class \emph{in a scope}, and bind it as a value
to a locally-scoped variable, which is then used to instantiate 
the created class in that scope. The |localClass| is a closure over
the mutable variable |x0|.

\begin{code}
 myNestedOOP =
   do
      localClass <- makeIncrementingPointClass
      localClass >>= ( # getX ) >>= Prelude.print
      localClass >>= ( # getX ) >>= Prelude.print
\end{code}

\begin{code}
 ghci> myNestedOOP
 1
 2
\end{code}

In contrast, such a class closure is not possible in Java, let alone
C++. Java supports anonymous objects, but not anonymous first-class
classes. Nested classes in Java must be linked to an object of the
enclosing class. Named nested classes in C\# are free from that
linking restriction. However, C\# does not support anonymous classes
or class computations in a local scope (although anonymous delegates
of C\# let us emulate computable classes). Nevertheless, classes, as
such, are not first-class citizens in any mainstream OO language.


\subsubsection{Open recursion}
\label{S:open-recursion}

The methods of an object may send messages to `self'. To support
inheritance with override that `self' must be bound
explicitly~\cite{CookThesis}. Otherwise, inheritance will not able to
revise the messages to self that were coded in a
superclass. Consequently, general object generators are to be given in
the style of `open recursion': they take self and construct (some part
of) self.

Quoting from~\cite{OCaml}[\S\,3.2]:

\begin{quote}\itshape\small
``A method or an initialiser can send messages to self (that is, the
current object). For that, self must be explicitly bound, here to the
variable @s@ (@s@ could be any identifier, even though we will often
choose the name @self@.) ... Dynamically, the variable @s@ is bound at
the invocation of a method. In particular, when the class
@printable_point@ is inherited, the variable @s@ will be correctly
bound to the object of the subclass.''
\end{quote}

\begin{code}
 class printable_point x_init =
   object (s)
     val mutable varX = x_init
     method getX      = varX
     method moveX d   = varX <- varX + d
     method print     = print_int s#getX
   end;;
\end{code}

Again, this OCaml code is transcribed to \OOHaskell\ very directly. The
self argument, @s@, ends up as an \emph{ordinary} argument of the
monadic function for generating printable point objects:

\begin{code}
 printable_point x_init s =
   do
      x <- newIORef x_init
      returnIO
        $  varX  .=. x
       .*. getX  .=. readIORef x
       .*. moveX .=. (\d -> modifyIORef x (+d))
       .*. print .=. ((s # getX ) >>= Prelude.print)
       .*. emptyRecord
\end{code}
%
In OCaml, we use |printable_point| as follows:

\begin{code}
 # let p = new printable_point 7;;
 val p : printable_point = <obj>
 # p#moveX 2;;
 - : unit = ()
 # p#print;;
 9- : unit = ()
\end{code}
Although @s@ does not appear on the line that constructs a point @p@
with the @new@ construct, the recursive knot clearly is tied right
there. In \OOHaskell, we use the (monadic) fixpoint function, @mfix@,
rather than a special keyword @new@. This makes the nature of openly
recursive object generators manifest.

\begin{code}
 mySelfishOOP =
   do
      p <- mfix (printable_point 7)
      p # moveX $ 2
      p # print
\end{code}

\begin{code}
 ghci> mySelfishOOP
 9
\end{code}


\subsubsection{Instantiation checking}

One potential issue with open recursion in \OOHaskell\ is that some
type errors in messages to self will not be spotted until the first
object construction is \emph{coded}. For instance, an OO library developer
could, accidentally, provide object generators that turn out to be
uninstantiatable; the library user would notice this defect once the generators
are put to work. This issue is readily resolved as follows. When we
program object generators, we may use the @concrete@ operation:

\begin{code}
 -- An assured printable point generator
 concrete_printable_point x_init 
   = concrete $ printable_point x_init
\end{code}

\begin{code}
 -- The concrete operation
 concrete generator self = generator self
  where
   _ = mfix generator
\end{code}
Operationally, @concrete@ is the identity function. However, it
constrains the type of @generator@ such that the application of @mfix@
is typeable. This approach needs to be slightly refined to cover
\emph{abstract} methods (aka pure virtual methods). To this end, one
would need to engage into local inheritance~---~adding vacuous (i.e.,
potentially undefined) methods for any needed virtual method. This
generalised @concrete@ operation would take the virtual portion of a
record, or preferably just a proxy for it, so that the purpose of this
argument is documented.


\subsection{Reuse techniques}

The first-class status of labels, methods and classes enables various,
common and advanced forms of reuse. Single inheritance boils down to
(monadic) function composition of object generators. Multiple
inheritance and object composition employ more advanced operations
of the record calculus.


\subsubsection{Single inheritance with extension}
\label{S:single-inheritance:extension}

Quoting from~\cite{OCaml}[\S\,3.7]:\footnote{We use British spelling
consistently in this paper, except for some words that enter the text through
code samples: color, colored, ...}

\begin{quote}\itshape\small
``We illustrate inheritance by defining a class of colored points that
inherits from the class of points. This class has all instance
variables and all methods of class @point@, plus a new instance
variable @color@, and a new method @getColor@.''
\end{quote}

\begin{code}
 class colored_point x (color : string) =
   object
     inherit point x
     val color = color
     method getColor = color
   end;;
\end{code}
Here is the corresponding OCaml session:

\begin{code}
 # let p' = new colored_point 5 "red";;
 val p' : colored_point = <obj>
 # p'#getX, p'#getColor;;
 - : int * string = (5, "red")
\end{code}

The following \OOHaskell\ version does not employ a special @inherit@
construct. We compose computations instead. To
construct a colored point we instantiate the superclass while
maintaining open recursion, and extend the intermediate record, @super@,
by the new method @getColor@.

\begin{code}
 colored_point x_init (color::String) self =
    do   super <- printable_point x_init self
         returnIO
             $  getColor .=. (returnIO color)
            .*. super
\end{code}
%
Here, @super@ is just a variable rather than an extra construct.

\begin{code}
 myColoredOOP =
   do
      p' <- mfix (colored_point 5 "red")
      x  <- p' # getX
      c  <- p' # getColor
      Prelude.print (x,c)
\end{code}
\OOHaskell\ and OCaml agree:

\begin{code}
 ghci>  myColoredOOP
 (5,"red")
\end{code}


\subsubsection{Class-polymorphic functionality}

We can parameterise computations with respect to \emph{classes}.

\begin{code}
 myFirstClassOOP point_class =
   do
      p <- mfix (point_class 7)
      p # moveX $ 35
      p # print
\end{code}

\begin{code}
 ghci> myFirstClassOOP printable_point
 42
\end{code}
The function @myFirstClassOOP@ takes a class (i.e., an object
generator) as an argument, instantiates the class, and moves and
prints the resulting object. We can pass @myFirstClassOOP@ any object
generator that creates an object with the slots @moveX@ and
@print@. This constraint is statically verified. For instance, the
colored point class, which we derived from the printable point class
in the previous section, is suitable:

\begin{code}
 ghci> myFirstClassOOP $ flip colored_point "red"
 42
\end{code}


\subsubsection{Single inheritance with override}
\label{S:single-inheritance:override}

We can \emph{override} methods and still refer to their superclass
implementations (akin to the @super@ construct in OCaml and other
languages). We illustrate overriding with a subclass of @colored_point@ whose
@print@ method is more informative:

\begin{code}
 colored_point' x_init (color::String) self =
    do
       super <- colored_point x_init color self
       return $  print .=. (
                   do  putStr "so far - "; super # print
                       putStr "color  - "; Prelude.print color )
             .<. super
\end{code}
%
The first step in the monadic @do@ sequence constructs an
old-fashioned colored point, and binds it to @super@ for further
reference. The second step in the monadic @do@ sequence returns
@super@ updated with the new @print@ method. The \HList\ operation
``@.<.@'' denotes type-preserving record update as opposed to the
familiar record extension ``@.*.@''. The operation ``@.<.@'' makes the
overriding explicit (as it is in C\#, for example). We could also use
a hybrid record operation, which does extension in case the given
label does not yet occur in the given record, falling back to
type-preserving update. This hybrid operation would let us model the
implicit overriding in C++ and Java. Again, such a variation point
demonstrates the programmability of \OOHaskell's object system.

Here is a demo that shows overriding to properly affect the @print@ method: 

\begin{code}
 myOverridingOOP =
   do
      p  <- mfix (colored_point' 5 "red")
      p  # print
\end{code}

\begin{code}
 ghci> myOverridingOOP
 so far - 5
 color  - "red"
\end{code}


\subsubsection{Orphan methods}

We can program methods outside of any hosting class. Such methods can
be reused across classes without any inheritance relationship. For
instance, we may define a method @print_getX@ that can be shared by
all objects that have at least the method @getX@ of the type
@Show@~@a@~@=>@~@IO@~@a@~---~regardless of any inheritance
relationships:

\begin{code}
 print_getX self = ((self # getX ) >>= Prelude.print)
\end{code}

We can update the earlier code for @printable_point@ as follows:

\begin{code}
 -- before: inlined definition of print
 ... .*. print    .=. ((s # getX ) >>= Prelude.print)
\end{code}

\begin{code}
 -- after: reusable orphan method
 ... .*. print    .=. print_getX s
\end{code}


\subsubsection{Flexible reuse schemes}

In addition to single class inheritance, there are several other established
reuse schemes in OO programming including object composition,
different forms of mixins and different forms of multiple
inheritance. Given the first-class status of all \OOHaskell\ entities
and its foundation in a powerful record calculus, it should be
possible to reconstruct most if not all existing reuse schemes. We
will use an (admittedly contrived) example to demonstrate a
challenging combination of multiple inheritance and object
composition. To the best of our knowledge, this example cannot be
directly represented in any existing mainstream language.


\begin{figure}[t]
\medskip
\begin{center}
\resizebox{.85\textwidth}{!}{\includegraphics{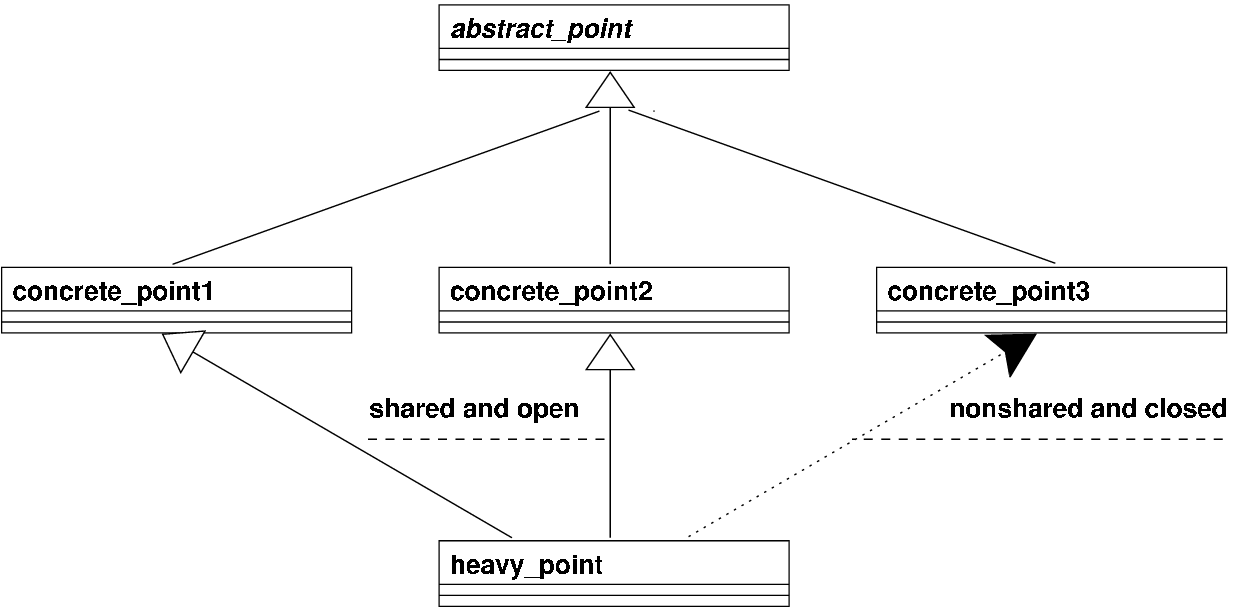}}
\end{center}
\caption{A complex reuse scenario}
\label{F:heavy}
\end{figure}


We are going to work through a scenario of making a class
@heavy_point@ from three different concrete subclasses of
@abstract_point@. The first two concrete points will be shared in the
resulting heavy point, because we leave open the recursive knot. The
third concrete point does not participate in the open recursion and is
not shared. In C++ terminology, |abstract_point| is a virtual base
class (with respect to the first two concrete points) and a
non-virtual base class at the same time. See Fig.~\ref{F:heavy} for an
overview.

The object template for heavy points starts as follows:

\begin{code}
 heavy_point x_init color self =
  do super1 <- concrete_point1 x_init self
     super2 <- concrete_point2 x_init self
     super3 <- mfix (concrete_point3 x_init)
     ... -- to be continued
\end{code}
We bind all ancestor objects for subsequent reference. We
pass @self@ to the first two points, which participate in open
recursion, but we fix the third point in place. The first two
classes are thus reused in the sense of inheritance, while the third class
is reused in the sense of object composition.  A heavy point
carries @print@ and @moveX@ methods delegating corresponding
messages to all three points:

\begin{code}
     ... -- continued from above
     let myprint = do
                      putStr "super1: "; (super1 # print)
                      putStr "super2: "; (super2 # print)
                      putStr "super3: "; (super3 # print)
     let mymove  = ( \d -> do
                              super1 # moveX $ d
                              super2 # moveX $ d
                              super3 # moveX $ d )
     return 
       $    print  .=. myprint
      .*.   moveX  .=. mymove
      .*.   emptyRecord
     ... -- to be continued
\end{code}

The three points, with all their many fields and methods, contribute
to the heavy point by means of left-biased union on records, which is
denoted by ``@.<++.@'' below:

\begin{code}
     ... -- continued from above
      .<++. super1
      .<++. super2
      .<++. super3
\end{code}

Here is a demo:

\begin{code}
 myDiamondOOP =
  do 
     p <- mfix (heavy_point 42 "blue")
     p # print -- All points still agree!
     p # moveX $ 2
     p # print -- The third point lacks behind!
\end{code}

\begin{code}
 ghci> myDiamondOOP
 super1: 42
 super2: 42
 super3: 42
 super1: 46
 super2: 46
 super3: 44
\end{code}

For comparison, in OCaml, multiple inheritance follows fixed rules.
Only the last definition of a method is kept: the redefinition in a
subclass of a method that was visible in the parent class overrides
the definition in the parent class. Previous definitions of a method
can be reused by binding the related ancestor using a special \ldots
@as@ \ldots\ notation. The bound name is said to be a `pseudo value
identifier' that can only be used to invoke an ancestor method. Eiffel,
C++, etc. have their own fixed rules and notations for multiple
inheritance. \OOHaskell\ allows us to ``program'' this aspect of the
OO type system. Programmers (or language designers) may devise their
own inheritance and object composition rules.


\subsection{Safe value recursion}
\label{S:mfix-safety}

The support for open recursion in an OO system has a
subtle but fundamental difficulty. Of the three ways to emulate open
recursion~--~recursive types, existential abstraction (\cite{ML-ART,
PT94}) and value recursion, the latter is the simplest one. 
This is the one we have chosen for \OOHaskell. Recall that each
object generator receives the @self@ argument (representing the constructed
object), which lets the methods send the messages
to the object itself. An object is constructed by obtaining the
fixpoint of the generator. Here is a variation on the printable point
example from Sec.~\ref{S:open-recursion} that illustrates the potential
unsafety of value recursion:

\begin{code}
 printable_point x_init self =
   do
      x <- newIORef x_init
      self # print  -- Unsafe!
      returnIO
        $  varX  .=. x
       .*. getX  .=. readIORef x
       .*. moveX .=. (\d -> modifyIORef x (+d))
       .*. print .=. ((self # getX ) >>= Prelude.print)
       .*. emptyRecord
\end{code}
%
An object generator may be tempted to invoke methods on the received
@self@ argument, as @self # print@ above. That code
typechecks. However, the attempt to construct an object by executing
 \[@mfix (printable_point 0)@\]
reveals the problem: looping. Indeed, @self@ represents the object
that \emph{will} be constructed. It is not proper to invoke any method
on @self@ when the object generation is still taking place, because
@self@ as a whole does not yet exist.

In Haskell, accessing a not-yet-constructed object leads to ``mere''
looping. This, so-called left-recursion problem (well-known in
parsing~\cite{ASU86}) is akin to the divergence of the following
trivial expression:

\[@mfix (\self -> do { Prelude.print self; return "s" })@\]

In a non-strict language like Haskell, determining the fixpoint of a
value @a->a@ where @a@ is not a function type is always safe: the
worst can happen is the divergence, but no undefined behaviour. In
strict languages, the problem is far more serious: accessing the field
before it was filled in is accessing a dummy value (e.g., a null
pointer) that was placed into the field prior to the evaluation of the
recursive definition. Such an access results in undefined behaviour,
and has to be prevented with a run-time check. As noted
in~\cite{ML-ART}, this problem has been widely discussed but no
satisfactory solution was found.

Although the problem is relatively benign in \OOHaskell\ and never leads
to undefined behaviour, we would like to statically prevent it. To be
precise, we impose the rule that the constructor may not execute any
actions that involve not-yet-constructed objects. With little changes,
we statically enforce that restriction. Object construction may be
regarded as a sort of a staged computation; the problem of preventing
the use of not-yet-constructed values is one of the key challenges
in multi-staged programming \cite{env-classifiers}, where it
has been recently solved with environment classifiers. Our solution is
related in principle (making the stage of completion of an object a
part of its type), but differs in technique (we exploit compile-time
tagging and monadic types rather than higher-ranked types).

We introduce a tag @NotFixed@ to mark the objects that are not
constructed yet:

\begin{code}
 newtype NotFixed a = NotFixed a -- data constructor opaque!
\end{code}
Because @NotFixed@ is a @newtype@, this tag imposes no run-time
overhead. We do not export the data constructor @NotFixed@ so the user
may not arbitrarily introduce or remove that tag. All operations on
this tag are restricted to a new module @NotFixed@ that is part of the
\OOHaskell\ library. The module exports two new operations: |new| and
|construct|. The former is a variant of @mfix@ for the @IO@ monad. The
@construct@ operation is a variation on @returnIO@. 

\begin{code}
 new :: (NotFixed (Record a)
          -> IO (NotFixed (Record a)))  -- object generator
     -> IO (Record a)                   -- object computation
 new f = mfix f >>= (\(NotFixed a) -> return a)
\end{code}

\begin{code}
 construct :: NotFixed (Record a)       -- self
           -> (Record a -> Record b)    -- constructor
           -> IO (NotFixed (Record b))  -- object computation
 construct (NotFixed self) f = returnIO $ NotFixed (f self)
\end{code}

Staged object construction proceeds as follows. The argument @self@
passed to the object generator is marked as @NotFixed@. After the
fixpoint is computed, @new@ removes the @NotFixed@ tag. The function
@construct@, while maintaining the @NotFixed@ tag, lifts the tag
internally so that the methods being defined by the object generator
could use the @self@ reference. We can now write our example as
follows:

\begin{code}
 printable_point x_init self =
   do
      x <- newIORef x_init
      -- self # print 
      construct self (\self->
           mutableX .=. x
       .*. getX     .=. readIORef x
       .*. moveX    .=. (\d -> modifyIORef x ((+) d))
       .*. print    .=. ((self # getX ) >>= Prelude.print)
       .*. emptyRecord)
\end{code}

\begin{code}
test_pp =
   do
      p <- new (printable_point 7)
      p # moveX $ 2
      p # print
\end{code}
%
If we uncomment the statement @self # print@ we will get the type
error saying that a @NotFixed@ object does not have the method
@print@. (There are no @HasField@ instances for the @NotFixed@ type.) 
Within the body of @construct@, the reference to @self@ is available
without the @NotFixed@ tag; so one may be tempted to invoke methods on
@self@ and execute their actions. However, the second argument of
@construct@ is a \emph{non-monadic} function of the type
@Record@~@a@~@->@~@Record@~@b@. Because the result type of the
function does not include @IO@, it is not possible to read and write
@IORef@ and do other @IO@ actions within that function. In Haskell (in
contrast to OCaml), imperativeness of a function is manifest in its
type.

The extension to the construction of inherited classes is
straightforward. For example, the @colored_point@ example from
Sec.~\ref{S:single-inheritance:extension} now reads:

\begin{code}
 colored_point x_init (color::String) self =
   do
        p <- printable_point x_init self
	-- p # print -- would not typecheck.
        construct p $ \p -> getColor .=. (returnIO color) .*. p
\end{code}

\begin{code}
 myColoredOOP =
   do
      p' <- new (colored_point 5 "red")
      x  <- p' # getX
      c  <- p' # getColor
      Prelude.print (x,c)
\end{code}
The constructor @colored_point@ receives the argument @self@ marked as
not-yet-constructed. We pass that argument to @printable_point@, which
gives us a not-yet-constructed object of the superclass. We cannot
execute any methods on that object (and indeed, uncommenting the
statement @p # print@ leads to a type error). The execution of a
superclass method may involve the invocation of a method on @self@
(as is the case for the method @print@), and @self@ is not constructed
yet. The @construct@ operation shown here is not fully general; the
source distribution illustrates safe object generation where methods
refer both to @self@ and @super@. The technique readily generalises to
multiple inheritance and object composition.


\section{Type-perceptive OOHaskell idioms}
\label{S:OOHaskell2}

So far we have avoided type declarations, type annotations and
explicit coercions of object types. We will now discuss those
\OOHaskell\ programming scenarios that can benefit from additional
type information, or even require it. We will pay special attention to
various subtyping and related cast and variance issues. In particular,
we will cover the technical details of subtype-polymorphic
collections, which require some amount of type perceptiveness, as we
saw in the introductory shapes example in Sec.~\ref{S:shapes}.


\subsection{Semi-implicit upcasting}
\label{S:lubNarrow}

There is an important difference between \OOHaskell's subtype
polymorphism (which we share to some extent with OCaml and ML-ART) and
polymorphism in C++ and other mainstream OO languages.\footnote{See,
  however, Sec.~\ref{S:nominal}, where we emulate the mainstream nominal
  subtyping.} In the latter languages, an object can be \emph{implicitly}
coerced to an object of any of its superclasses (``upcast''). One may
even think that an object is polymorphic by itself, i.e., it has types
of all of its superclasses, simultaneously. Hence, there is no need for
functions on objects (or methods) to be polymorphic by themselves; they
are monomorphic.

In OCaml and \OOHaskell, it is the other way around: objects are
monomorphic (with regard to the record structure) and the language
semantics does not offer any implicit upcasting or \emph{narrowing}.\footnote{We prefer the term
  \emph{narrow} over \emph{up-cast}, as to emphasise the act of
  restricting the interface of an object, as opposed to walking up an
  explicit (perhaps even nominal) subtyping hierarchy.}  However,
functions that take objects can be polymorphic and can process objects
of different types. To be precise, \OOHaskell\ exploits type-class-bounded polymorphism. A function that takes an object and refers to
its methods (i.e., record components) has in its inferred or explicit
type @HasField@ constraints for these record components. The function
therefore accepts \emph{any} object that has \emph{at least} the
components that satisfy the @HasField@ constraints.\footnote{We
  oversimplify here by not talking about operations that add or remove fields. This is a fair simplification though because we talk about normal OO functionality here, as opposed to free-wheeling functionality for
record manipulation.}
Therefore, most of the time no (implicit or explicit) upcasting is
needed; in fact, in Sec.~\ref{S:OOHaskell1} we did not see
any.

An explicit cast is usually understood as casting to an explicitly
named target type. We discuss such casts later in this section. Here
we show that the established explicit-vs.-implicit upcast dichotomy
misses an intermediate option, which is admitted by
\OOHaskell. Namely, \OOHaskell\ lets the programmer specify that
narrowing is to be performed~---~without giving an explicit target type, though. So we
continue to get by without specifying types, leaving it all to type
inference~---~at least for a while.

In \OOHaskell\ (and OCaml), we must narrow an object if its expression
context has no type-class-bounded polymorphism left and requires an
object of a different type. The archetypal example is placing objects
in a homogeneous collection, e.g., a list. The original item objects
may be of different types; therefore, we must establish a common
element type and narrow the items to it. This common element type does
not have to be specified explicitly, however. \OOHaskell\ can compute
that common type as we add objects to the collection; the context will
drive the narrowing. The \OOHaskell\ implementation of the shapes
example in Sec.~\ref{S:shapes} involved this sort of narrowing:

\begin{code}
 myOOP = do
            s1 <- mfix (rectangle (10::Int) (20::Int) 5 6)
            s2 <- mfix (circle (15::Int) 25 8)
            let scribble = consLub s1 (consLub s2 nilLub)
            ... and so on ...
\end{code}
The designated list constructors @nilLub@ and @consLub@ incorporate
narrowing into their normal constructor behaviour. The specific element
type of each new element constraints the ultimate least-upper bound
(LUB) element type for the final list. Elements are continuously cast
towards this LUB. The list constructors are defined as follows:

\begin{Verbatim}[fontsize=\small,commandchars=\\\{\}]
 -- A type-level code for the empty list
 data NilLub
\end{Verbatim}

\begin{Verbatim}[fontsize=\small,commandchars=\\\{\}]
 -- The empty list constructor
 nilLub = \undefined :: NilLub
\end{Verbatim}

\begin{Verbatim}[fontsize=\small,commandchars=\\\{\}]
 -- Cons as a type-level function
 class ConsLub h t l | h t -> l
  where
   consLub :: h -> t -> l
\end{Verbatim}

\begin{Verbatim}[fontsize=\small,commandchars=\\\{\}]
 -- No coercion needed for a singleton list
 instance ConsLub e  NilLub [e]
  where
   consLub h _ = [h]
\end{Verbatim}

\begin{Verbatim}[fontsize=\small,commandchars=\\\{\}]
 -- Narrow head and tail to their LUB type
 instance LubNarrow e0 e1 e2 => ConsLub e0 [e1] [e2]
  where
   consLub h t = fst (head z) : map snd (tail z)
    where
     z = map (lubNarrow h) (\undefined:t)
\end{Verbatim}

The important operation is @lubNarrow@:

\begin{code}
 class  LubNarrow a b c | a b -> c
  where lubNarrow :: a -> b -> (c,c)
\end{code}
Given two values of record types @a@ and @b@, this operation returns a
pair of narrowed values, both of type @c@, where @c@ is supposed to be
the least-upper bound in the sense of structural subtyping. The
specification of @lubNarrow@ once again illustrates the capability of
\OOHaskell\ to `program' OO type-system aspects. We exploit the
type-level reflection on \HList\ records to define narrowing:

\begin{code}
 instance ( HZip la va a
          , HZip lb vb b
          , HTIntersect la lb lc
          , H2ProjectByLabels lc a c aout
          , H2ProjectByLabels lc b c bout
          , HRLabelSet c
          )
       => LubNarrow (Record a) (Record b) (Record c)
  where
   lubNarrow ra@(Record a) rb@(Record b) =
      ( hProjectByLabels lc ra
      , hProjectByLabels lc rb
      )
    where
     lc = hTIntersect la lb
     (la,_) = hUnzip a
     (lb,_) = hUnzip b
\end{code}
That is, given two records @ra@ and @rb@, we compute the intersection @lc@ of
their labels @la@ and @lb@ such that we can subsequently project both
records to this shared label set. It is possible to improve @consLub@
so that we can construct lists in linear time. We may also want to
consider depth subtyping in addition to width subtyping, as we will
discuss in Sec.~\ref{S:deep}.

\subsection{Narrow to a fixed type}
\label{S:narrow}

The LUB narrowing is neither an explicit nor an implicit coercion. In
the shapes example, we explicitly apply special list constructors,
which we know perform coercion, but the target type is left
implicit. Such semi-implicit narrowing is a feature of \OOHaskell, not
available in the otherwise similar systems OCaml and ML-ART. In OCaml,
building the @scribble@ list in the shapes example requires fully
explicit narrowing (which OCaml calls upcast, ``@:>@''):

\begin{code}
   let (scribble: shape list) = [
      (new rectangle 10 20 5 6 :> shape);
      (new circle 15 25 8 :> shape)] in ...
\end{code}
We can express such narrowing in \OOHaskell\ as well:

\begin{code}
   s1 <- mfix (rectangle (10::Int) (20::Int) 5 6)
   s2 <- mfix (circle (15::Int) 25 8)
   let scribble :: [Shape Int]
   scribble = [narrow s1, narrow s2]
\end{code}
The applications of @narrow@ prepare the shape objects for insertion into
the homogeneous Haskell list. We do not need to identify the
target type per element: specifying the desired type for the result
list is enough. The operation @narrow@ is defined in a dedicated class:

\begin{code}
 class  Narrow a b
  where  narrow :: Record a -> Record b
\end{code}
The operation @narrow@ extracts those label-value pairs from @a@ that are requested by
@b@. Its implementation uses the same kind of projection on
records that we saw in full in the previous section; cf.\ @lubNarrow@.

(Fully) explicit narrowing implies that we must declare appropriate
types~---~something that we managed to avoid so far. Here is the
@Shape@ type, which `drives' narrowing in the example:

\begin{code}
 -- The Shape interface
 type Shape a = Record (  GetX    :=: IO a
                      :*: GetY    :=: IO a
                      :*: SetX    :=: (a -> IO ())
                      :*: SetY    :=: (a -> IO ())
                      :*: MoveTo  :=: (a -> a -> IO ())
                      :*: RMoveTo :=: (a -> a -> IO ())
                      :*: Draw    :=: IO ()
                      :*: HNil )
\end{code}
Two infix type synonyms add convenience to explicitly written types:

\begin{code}
 infixr 2 :*:
 type e :*: l = HCons e l
\end{code}

\begin{code}
 infixr 4 :=:
 type l :=: v = (l,v)
\end{code}
The @Shape@ interface above explicitly includes the virtual operation
@draw@ because the loop over @scribble@ needs this method.  We will
see more applications of @narrow@ in subsequent sections.


\subsection{Self-returning methods}
\label{S:self}

{\sloppypar

A self-returning method is a method whose result type is the type of
self or is based on it. An example is a @clone@ method. The
typing of such methods (and of @self@) is known to be a difficult
issue in typed object encodings; cf.~\cite{CHC90,AC96,PolyTOIL} for
some advanced treatments. In \OOHaskell, we must not naively define
a method @me@ that returns @self@, as is:

}

\begin{code}
 self_returning_point (x_init::a) self =
   do
      super <- printable_point x_init self
      returnIO
          $  me .=. self -- WRONG!
         .*. super
\end{code}
If we wrote such code, then we get a type error, when we attempt to instantiate the
corresponding object (i.e., when we @mfix@ the object generator).
Haskell does not permit (equi-)recursive types, which are needed to
type @self@ in the example. The issue of recursive types and returning
the full @self@ is discussed in detail in
Sec.~\ref{S:recursive}. Here, we point out a simpler solution:
disallowing returning @self@ and requiring the programmer to narrow
@self@ to a specific desired interface. In the case of the clone
method, mainstream programming languages typically define its
return type to be the base class of all classes. The programmer is
then supposed to use downcast to the intended subtype.

We resolve the problem with the self-returning method as follows:

\begin{code}
 self_returning_point (x_init::a) self =
   do
      super <- printable_point x_init self
      returnIO
          $  me .=. (narrow self :: PPInterface a)
         .*. super
\end{code}

\begin{code}
 type PPInterface a
    = Record (  GetX  :=: IO a
            :*: MoveX :=: (a -> IO ())
            :*: Print :=: IO ()
            :*: HNil )
\end{code}
That is, @me@ narrows @self@ explicitly to the interface for printable points.

We should relate the explicit narrowing of the return type of @me@ to
the explicit declaration of the return type of all methods in C++ and
Java. The presented @narrowing@ approach does have a limitation
however: all record components that do not occur in the target
interface are irreversibly eliminated from the result record.  We
would prefer to make these components merely `private' so they can
be recovered through a safe downcast. We offer two options for such
downcastable upcasts in the next two sections.


\subsection{Casts based on dynamics}

Turning again to the shapes benchmark, let us modify the loop over
@scribble@, a homogeneous list of shapes, so to single out circles for
special treatment. This requires downcast:

\begin{code}
       mapM_ (\shape -> maybe (putStrLn "Not a circle.")
                              (\circ -> do circ # setRadius $ 10;
                                           circ # draw)
                              ((downCast shape) `asTypeOf` (Just s2)))
             scribble
\end{code}
%
In each iteration, we attempt to downcast the given shape
object to the type of @s2@ (which we recall is a circle object).  A
downcast may fail or succeed, hence the |Maybe| type of the result.

{\sloppypar

Neither \OOHaskell's @narrow@ operation nor OCaml's upcast support
such scenarios. \OOHaskell's @narrow@ irrevocably removes record
components. We can define, however, other forms of upcast, which are
reversible. We begin with a technique that exploits dynamic
typing~\cite{ACPP89,ACPP91,LPJ03}.

}

The new scribble list is built as follows:

\begin{code}
 let scribble :: [UpCast (Shape Int)]
     scribble = [upCast s1, upCast s2]
\end{code}
where
\begin{code}
 data UpCast x = UpCast x Dynamic
\end{code}
The data constructor |UpCast| is opaque for the library user, who can
only up-cast through a dedicated @upCast@ operation. The latter saves
the original object by embedding it into @Dynamic@. (We presume that
record types readily instantiate the type class @Typeable@.) Dually,
downcast is then a projection from @Dynamic@ to the requested subtype:

\begin{code}
 upCast :: (Typeable (Record a), Narrow a b)
        => Record a -> UpCast (Record b)
 upCast x = UpCast (narrow x) (toDyn x)
\end{code}

\begin{code}
 downCast :: (Typeable b, Narrow b a)
          => UpCast (Record a) -> Maybe (Record b)
 downCast (UpCast _ d) = fromDynamic d
\end{code}

We want to treat `upcast objects' as being objects too, and so we add
a trivial @HasField@ instance for looking up record components of
upcast objects. This instance delegates the look-up to the narrowed
part of the @UpCast@ value:

\begin{code}
 instance HasField l x v => HasField l (UpCast x) v
  where
   hLookupByLabel l (UpCast x _) =  hLookupByLabel l x
\end{code}

This technique suffers from a few shortcomings. Although downcast is
safe in a sense that no `bad things can happen' (cf.\ unsafe casts in
C), this downcast does not keep us from attempting so-called `stupid
casts', i.e., casts to types for which casting cannot possibly
succeed. In the following section, we describe a more elaborate
upcast/downcast pair that statically prevents stupid downcasts. The
dynamics-based method also suffers from the full computational
overhead of the narrow operation, a value-level coercion that iterates
over all record components.


\subsection{Casts based on unions}
\label{S:union}

We turn to the subtyping technique from Sec.~\ref{S:mutable} (further
refined in Sec.~\ref{S:objcomp}), which used union types to represent
the intersection of types. That techniques had several problems:
it could not easily deal with the empty list, could not minimise the
union type to the number of distinct element types, and could not
downcast. We fully lift these restrictions here by putting
type-level programming to work.

We again make upcasts semi-implicit with dedicated list constructors:

\begin{code}
 myOOP = do
            s1 <- mfix (rectangle (10::Int) (20::Int) 5 6)
            s2 <- mfix (circle (15::Int) 25 8)
            let scribble = consEither s1 (consEither s2 nilEither)
            ... and so on ...
\end{code}
The list constructors are almost identical to @nilLub@ and @consLub@ in
Sec.~\ref{S:lubNarrow}. The difference comes when we cons to a
non-empty list; see the last instance below:

\begin{code}
 -- A type-level code for the empty list
 data NilEither
\end{code}

\begin{Verbatim}[fontsize=\small,commandchars=\\\{\}]
 -- The empty list constructor
 nilEither = \undefined :: NilEither
\end{Verbatim}

\begin{code}
 -- Cons as a trivial type-level function
 class ConsEither h t l | h t -> l
  where
   consEither :: h -> t -> l
\end{code}

\begin{code}
 -- No coercion needed for a singleton list
 instance ConsEither e  NilEither [e]
  where
   consEither h _ = [h]
\end{code}

\begin{code}
 -- Construct union type for head and tail
 instance ConsEither e1 [e2] [Either e1 e2]
  where
   consEither h t = Left h : map Right t
\end{code}
We extend the union type for the ultimate element type with one branch
for each new element, just as we did in the Haskell~98-based encoding
of Sec.~\ref{S:objcomp}. However, with type-level programming,
we can, \emph{in principle}, minimise the union type so that each
distinct element type occurs exactly once.\footnote{The same kind of
  constraints was covered in the \HList\ paper~\cite{HLIST-HW04},
  cf.~\emph{type-indexed} heterogeneous collections.} This
straightforward optimisation is omitted here for brevity.\footnote{In
  essence, we need to iterate over the existing union type and use
  type-level type equality to detect if the type of the element to
  cons has already occurred in the union. If so, we also need to
  determine the corresponding sequence of @Left@ and @Right@ tags.}

Method look-up is generic, treating the union type as the intersection
of record fields of the union branches:

\begin{code}
 instance (HasField l x v, HasField l y v) 
        => HasField l (Either x y) v 
  where
   hLookupByLabel l (Left x)  =  hLookupByLabel l x
   hLookupByLabel l (Right y) =  hLookupByLabel l y
\end{code}

Downcast is a type-driven search operation on the union type. We also
want downcast to fail statically if the target types does not appear
among the branches. Hence, we start downcast with a type-level
Boolean, @hFalse@, to express that we have not yet seen the type in
question:

\begin{Verbatim}[fontsize=\small,commandchars=\|\{\}]
 downCast = downCastSeen hFalse
\end{Verbatim}
Downcast returns |Maybe| because it can intrinsically fail at the value level:

\begin{Verbatim}[fontsize=\small,commandchars=\|\{\}]
 class  DownCastSeen seen u s
  where downCastSeen :: seen -> u -> Maybe s
\end{Verbatim}

We process the union like a list. Hence, there are two cases: one for
the non-singleton union, and one for the final branch. Indeed, the 
details of the definition reveal that we assume right-associative unions.

\begin{Verbatim}[fontsize=\small,commandchars=\|\{\}]
 instance (DownCastEither seen b x y s, TypeEq x s b) 
       =>  DownCastSeen seen (Either x y) s
  where
    downCastSeen seen = downCastEither seen (|undefined::b)
\end{Verbatim}

\begin{Verbatim}[fontsize=\small,commandchars=\|\{\}]
 instance (TypeCastSeen seen b x s, TypeEq x s b)
       =>  DownCastSeen seen x s
  where
    downCastSeen seen = typeCastSeen seen (|undefined::b)
\end{Verbatim}
In both cases we test for the type equality between the target type
and the (left) branch type. We pass the computed type-level
Boolean to type-level functions @DownCastEither@ (`non-singleton
union') and @TypeCastSeen@ (`final branch', a singleton union):

\begin{code}
 class  TypeCastSeen seen b x y 
  where typeCastSeen :: seen -> b -> x -> Maybe y
\end{code}

\begin{code}
 instance TypeCast x y => TypeCastSeen seen HTrue x y
  where   typeCastSeen _ _ = Just . typeCast
\end{code}

\begin{code}
 instance TypeCastSeen HTrue HFalse x y
  where   typeCastSeen _ _ = const Nothing
\end{code}
The first instance applies when we have encountered the requested type
at last. In that case, we invoke normal, type-level type cast
(cf.~\cite{HLIST-HW04}), knowing that it must succeed given the
earlier check for type equality. The second instance applies when the
final branch is not of the requested type. However, we must have seen
the target type among the branches, cf.~@HTrue@. Thereby, we rule out
stupid casts.

The following type-level function handles `non-trivial' unions:
\begin{code}
 class  DownCastEither seen b x y s
  where downCastEither :: seen -> b -> Either x y -> Maybe s
\end{code}

\begin{code}
 instance (DownCastSeen HTrue y s, TypeCast x s)
       =>  DownCastEither seen HTrue x y s
  where
    downCastEither _ _ (Left x)  = Just (typeCast x)
    downCastEither _ _ (Right y) = downCastSeen hTrue y
\end{code}

\begin{code}
 instance DownCastSeen seen y s
       => DownCastEither seen HFalse x y s
  where
    downCastEither _ _    (Left x)  = Nothing
    downCastEither seen _ (Right y) = downCastSeen seen y
\end{code}
The first instances applies in case the \emph{left} branch of the
union type is of the target type; cf.\ @HTrue@. It remains to check
the value-level tag. If it is |Left|, we are done after the type-level
type cast. We continue the search otherwise, with |seen| set to
|HTrue| to record that the union type does indeed contain the target
type. The second instance applies in case the left branch of the
union type is \emph{not} of the target type; cf.\ @HFalse@. In that
case, downcast continues with the tail of union type, while
propagating the |seen| flag. Thereby, we rule out stupid casts.


\subsection{Explicit type constraints}
\label{S:constrain}

In some cases, it is useful to impose structural record type
constraints on arguments of an object generator, on arguments or the
result type of a method. These constraints are akin to C++
\emph{concepts}~\cite{siek05:_concepts_cpp0x}. The familiar @narrow@
turns out to be a convenient tool for imposition of such type constraints.
This use of @narrow@ does no operations at run-time. A good example
example of OO type constraints is the treatment of virtual methods in
\OOHaskell.

Quoting from~\cite{OCaml}[\S\,3.4]:

\begin{quote}\itshape\small
``It is possible to declare a method without actually defining it,
using the keyword @virtual@. This method will be provided later in
subclasses. A class containing virtual methods must be flagged
virtual, and cannot be instantiated (that is, no object of this class
can be created). It still defines type abbreviations (treating virtual
methods as other methods.)
\end{quote}

\begin{code}
 class virtual abstract_point x_init =
   object (self)
     val mutable varX = x_init
     method print = print_int self#getX
     method virtual getX : int
     method virtual moveX : int -> unit
   end;;
\end{code}

In C++, such methods are called \emph{pure} virtual and the
corresponding classes are called abstract. In Java and C\#, we can
flag both methods and classes as being abstract. In \OOHaskell, it is
enough to leave the method undefined. Indeed, in the shapes example, we
omitted any mentioning of the @draw@ method when we defined the object
generator for shapes.

OCaml's abstract point class may be transcribed to \OOHaskell\ as follows:

\begin{code}
 abstract_point x_init self =
   do
      xRef <- newIORef x_init
      returnIO $
           varX   .=. xRef
       .*. print  .=. ( self # getX >>= Prelude.print )
       .*. emptyRecord
\end{code}
%
This object generator cannot be instantiated with @mfix@ because
@getX@ is used but not defined. The Haskell type system effectively
prevents us from instantiating classes which use the methods neither
they nor their parents have defined. There arises the question of the
explicit designation of a method as pure virtual, which would be of
particular value in case the pure virtual does not happen to be used
in the object generator itself.

\OOHaskell\ allows for such explicit designation by means of adding
type constraints to @self@. To designate @getX@ and @moveX@ as pure
virtuals of @abstract_point@ we change the object generator as
follows:

\begin{code}
 abstract_point (x_init::a) self =
   do
      ... as before ...
 where
  _ = narrow self :: Record (  GetX  :=: IO a
                           :*: MoveX :=: (a -> IO ())
                           :*: HNil )
\end{code}
We use the familiar @narrow@ operation, this time to express a type
constraint. We must stress that we narrow here at the type level
only. The result of narrowing is not used (cf.\ ``\_''), so
operationally it is a no-op. It does however affect the typechecking
of the program: every instantiatable extension of @abstract_point@
must define @getX@ and @moveX@. 

One may think that the same effect can be achieved by adding regular
type annotations (e.g., on @self@). These annotations however must
spell out the desired object type entirely. Furthermore, a regular
record type annotation rigidly and unnecessarily restrains the order
of the methods in the record as well as their types (preventing deep
subtyping, Sec.~\ref{S:deep}). One may also think object types can be
simply constrained by specifying @HasField@ constraints. This is
impractical in so far that \emph{full} object types would need to be
specified then by the programmer; Haskell does not directly support
partial signatures. Our @narrow@-based approach solves these problems.


\subsection{Nominal subtyping}
\label{S:nominal}

In OCaml and, by default, in \OOHaskell, object types engage into
structural subtype polymorphism. Many other OO languages prefer
nominal object types with explicitly declared subtyping (inheritance)
relationships. There is an enduring debate about the superiority of
either form of subtyping. The definite strength of structural subtype
polymorphism is that it naturally enables inference of object
types. The downside is potentially accidental subtyping~\cite{CW85}:
a given object may be admitted as an actual argument of some function
just because its structural type fits. Nominal types allow us to
restrict subtyping polymorphism on the basis of explicitly declared
subclass or inheritance relationships between nominal (i.e., named)
types.

Although \OOHaskell\ is biased towards structural subtyping
polymorphism, \OOHaskell, as a general sandbox for typed OO language
design, does admit nominal object types and nominal subtyping
including multiple inheritance.

We revisit our familiar printable points and colored points, 
switching to nominal types. First, we need to invent class names,
or nominations:

\begin{code}
 data PP  = PP  -- Printable points
 data CP  = CP  -- Colored points
\end{code}
As an act of discipline, we also register these types as nominations:

\begin{code}
 class Nomination f
\end{code}

\begin{code}
 instance Nomination PP
 instance Nomination CP
\end{code}

We attach nomination to a regular, record-based \OOHaskell\ object as a
phantom type. To this end, we using the following @newtype@ wrapper:
\begin{code}
 newtype N nom rec = N rec
\end{code}
The following two functions add and remove the nominations:

\begin{code}
 -- An operation to `nominate' a record as nominal object
 nominate ::  Nomination nt => nt -> x -> N nt x
 nominate nt x = N x
\end{code}

\begin{code}
 -- An operation to take away the type distinction
 anonymize ::  Nomination nt => N nt x -> x
 anonymize (N x) = x
\end{code}
To be able to invoke methods on nominal objects, we need a @HasField@
instance for @N@, with the often seen delegation to the wrapped
record:

\begin{code}
 instance (HasField l x v, Nomination f) => HasField l (N f x) v
  where hLookupByLabel l o = hLookupByLabel l (anonymize o)
\end{code}

OO programming with nominal subtyping on @PP@ and @CP@ can now
commence. The object generator for printable points remains exactly
the same as before except that we nominate the returned object as an
@PP@:

\begin{code}
 printable_point x_init s =
   do
       x <- newIORef x_init
       returnIO $ nominate PP -- Nominal!
         $  mutableX .=. x
        .*. getX     .=. readIORef x
        .*. moveX     .=. (\d -> modifyIORef x (+d))
        .*. print    .=. ((s # getX ) >>= Prelude.print)
        .*. emptyRecord
\end{code}

The nominal vs.~structural distinction only becomes meaningful once we
start to annotate functions explicitly with the requested
nominal argument type. We will first consider request that
insist on a specific nominal type, with no subtyping involved.
Here is a print function that only accepts nominal printable points.

\begin{code}
 printPP (aPP::N PP x) = aPP # print
\end{code}

To demonstrate nominal subtyping, we define colored points (`@CP@'):

\begin{code}
 colored_point x_init (color::String) self =
    do
       super <- printable_point x_init self
       returnIO $ nominate CP -- Nominal!
         $  print .=. ( do  putStr "so far - "; super # print
                            putStr "color  - "; Prelude.print color )
        .<. getColor .=. (returnIO color)
        .*. anonymize super -- Access record!
\end{code}
We need to make @CP@ a nominal subtype of @PP@. That designation is
going to be explicit. We introduce a type class @Parents@,
which is an extensible type-level function from nominal types to the
list of their immediate supertypes. A type may have more than one
parent: multiple inheritance. The following two instances designate
@PP@ as the root of the hierarchy and @CP@ as its immediate subtype:

\begin{code}
 class ( Nomination child, Nominations parents ) =>
       Parents child parents | child -> parents
\end{code}

\begin{code}
 instance Parents PP HNil             -- PP has no parents
 instance Parents CP (HCons PP HNil)  -- Colored points are printable points
\end{code}
The \OOHaskell\ library also defines a general relation @Ancestor@, which
is the reflexive, transitive closure of @Parents@:

\begin{code}
 class ( Nomination f, Nomination anc ) =>
       Ancestor f anc
\end{code}
We are now in the position to define an upcast operation, which is the basis for
nominal subtyping:

\begin{code}
 -- An up-cast operation
 nUpCast :: Ancestor f g => N f x -> N g x
 nUpCast = N . anonymize
\end{code}
We could also define some forms of downcast. Our @nUpCast@ does no
narrowing, so operationally it is the identity function. This is
consistent with the implementation of the nominal upcast in mainstream
OO languages. The record type of an \OOHaskell\ object is still
visible in its nominal type. Our nominal objects are fully \OOHaskell\
objects except that their subtyping is deliberately restricted. 

We can define a subtype-polymorphic print function for printable
points by `relaxing' the non-polymorphic @printPP@ function through
upcast.\footnote{We cannot define @printPP'@ in a point-free style
because of Haskell's monomorphism restriction.}

\begin{code}
 printPP (aPP::N PP x) = aPP # print -- accept PP only
 printPP' o = printPP (nUpCast o)    -- accept PP and nominal subtypes
\end{code}

The couple @printPP@ and @printPP'@ clarifies that we can readily
restrict argument types of functions to either precise types or to all
subtypes of a given base. This granularity of type constraints it not
provided by mainstream OO languages. Also, the use of structural
subtyping in the body of @printPP@ hints at the fact that we can blend
nominal and structural subtyping with ease in \OOHaskell. Again,
this is beyond state-of-the-art in mainstream OO programming.


\subsection{Iso-recursive types}
\label{S:recursive}

In the previous section, we have studied nominal types for the sake of
nominal subtyping. Nominal types are intrinsically necessary, when we
need to model recursive object types in \OOHaskell. In principle, a
type system with equi-recursive types would be convenient in this
respect. However, adding such types to Haskell was debated and then
rejected because it will make type-error messages nearly
useless~\cite{Hughes02}. Consequently, we encode recursive object
types as iso-recursive types; in fact, we use @newtype@s. (An
alternative technique of existential quantification~\cite{PT94} is
discussed in Sec.~\ref{S:anti-patterns}.)

We illustrate iso-recursive types on uni-directionally linked dynamic
lists. The interface of such list objects has methods that also return
list objects: a getter for the tail and an insertion method.

\begin{code}
 -- The nominal object type
 newtype ListObj a =
         ListObj (ListInterface a)
\end{code}

\begin{code}
 -- The structural interface type
 type ListInterface a =
      Record (     IsEmpty :=: IO Bool
               :*: GetHead :=: IO a
               :*: GetTail :=: IO (ListObj a)
               :*: SetHead :=: (a -> IO ())
               :*: InsHead :=: (a -> IO (ListObj a))
               :*: HNil )
\end{code}
Recall that we had to define a @HasField@ instance whenever we went
beyond the normal `objects as records' approach.  This is the case
here, too. Each newtype for iso-recursion has to be complemented by a
trivial @HasField@ instance:

\begin{code}
 instance HasField l (ListInterface a) v =>
          HasField l (ListObj a) v
   where
   hLookupByLabel l (ListObj x) = hLookupByLabel l x
\end{code}

For clarity, we chose the implementation of @ListInterface a@ with two
OO classes: for the empty and non-empty lists. A single OO list class
would have sufficed too. Empty-list objects fail for all
getters. Here is the straightforward generator for empty lists:

\begin{code}
 nilOO self :: IO (ListInterface a)
  = returnIO
     $  isEmpty  .=. returnIO True
    .*. getHead  .=. failIO "No head!"
    .*. getTail  .=. failIO "No tail!"
    .*. setHead  .=. const (failIO "No head!")
    .*. insHead  .=. reusableInsHead self
    .*. emptyRecord
\end{code}
%
The reusable insert operation constructs a new object of the @consOO@:

\begin{code}
 reusableInsHead list head
  = do 
       newCons <- mfix (consOO head list)
       returnIO (ListObj newCons)
\end{code}
Non-empty list objects hold a reference for the head, which is
accessed by @getHead@ and @setHead@. Here is the object generator
for non-empty lists:

\begin{code}
 consOO head tail self
  = do
       hRef <- newIORef head
       returnIO
         $  isEmpty .=. returnIO False
        .*. getHead .=. readIORef hRef
        .*. getTail .=. returnIO (ListObj tail)
        .*. setHead .=. writeIORef hRef
        .*. insHead .=. reusableInsHead self
        .*. emptyRecord
\end{code}
OO programming on nominal objects commences without ado. They can be used
just like record-based \OOHaskell\ objects before. As an example,
the following recursive function prints a given list. One can check
that the various method invocations involve nominally typed objects.

\begin{code}
 printList aList
  = do
       empty <- aList # isEmpty
       if empty
         then putStrLn ""
         else do 
                 head <- aList # getHead
                 putStr $ show head
                 tail <- aList # getTail
                 putStr " "
                 printList tail
\end{code}


\subsection{Width and depth subtyping}
\label{S:deep}

We have used the term subtyping in the informal sense of type-safe
type substitutability. That is, we call the object type $S$ to be a
subtype of the object type $T$ if in any well-typed program $P$ the
typeability of method invocations is preserved upon replacing objects
of type $T$ with objects of type $S$. This notion of subtyping is to
be distinguished from behavioural subtyping, also known as Liskov
Substitution Principle~\cite{LiskovW94}.

In \OOHaskell, subtyping is enabled by the type of the method
invocation operator @#@. For instance, the function @\o -> o # getX@
has the following inferred type:
\[@HasField (Proxy GetX) o v => o -> v@\]
This type is polymorphic. The function will accept any object (i.e.,
record) @o@ provided that it has the method labelled @GetX@ whose type
matches the function's desired return type @v@.

A basic form of subtyping or subsumption is width subtyping, whereupon
an object of type $S$ is a subtype of $T$ if the record type $S$ has
(at least) all the fields of $T$ with the exact same type. The \HList\
library readily provides this subtyping relation,
@Record.SubType@. Corresponding constraints can be added to type
signatures (although we recall that Sec.~\ref{S:constrain} devised a
constraint technique that is more convenient for \OOHaskell). It is
easy to see that if @SubType@~$S$~$T$ holds for some record types $S$
and $T$, then substituting an object of type $S$ for an object of type
$T$ preserves the typing of every occurrence of @#@ in a program. No
method will be missing and no method will be of a wrong type.

Width subtyping is only one form of subtyping. There are other
subtyping relations, which too preserve the typing of each occurrence
of @#@ in a program~---~in particular, depth subtyping. While width
subtyping allows the subtype to have more fields than the supertype,
depth subtyping allows the fields of the subtype to relate to the
fields of the supertype by subtyping.  Typed mainstream OO languages
like Java and C\# do not support full depth subtyping.

We will now explore depth subtyping in \OOHaskell. We define some new
object types and functions on the one-dimensional @printable_point@
class from Sec.~\ref{S:open-recursion} and its extension
@colored_point@ from Sec.~\ref{S:single-inheritance:override}. We
define a simple-minded one-dimensional vector class, specified by two
points for the beginning and the end, which can be accessed by the
methods @getP1@ and @getP2@:

\begin{code}
 vector (p1::p) (p2::p) self =
   do
      p1r <- newIORef p1
      p2r <- newIORef p2
      returnIO $
           getP1    .=. readIORef p1r
       .*. getP2    .=. readIORef p2r
       .*. print    .=. do self # getP1 >>= ( # print )
                           self # getP2 >>= ( # print )
       .*. emptyRecord
\end{code}
%
The local type annotations @p1::p@ and @p2::p@ enforce our intent that
the two points of the vector have the same type. It is clear that
objects of type @p@ must be able to respond to the message
@print@. Otherwise, the type of the points is not constrained. Our
object generator @vector@ is parameterised over the class of
points. In C++, the close analogue is a class \emph{template}. This
example shows that Haskell's normal forms of polymorphism, combined
with type inference, allow us to define parameterised classes without
ado.

We construct two vector objects, @v@ and @cv@:

\begin{code}
 testVector = do
                 p1  <- mfix (printable_point 0)
                 p2  <- mfix (printable_point 5)
                 cp1 <- mfix (colored_point 10 "red")
                 cp2 <- mfix (colored_point 25 "red")
                 v  <- mfix (vector p1 p2)
                 cv <- mfix (vector cp1 cp2)
                 -- ... to be continued ...
\end{code}
The former is the vector of two printable points; the latter is the
vector of two colored points. The types of @v@ and @cv@ are obviously
different: the type checker will remind us of this fact if we tried to
put both vectors into the same homogeneous list. The vectors @v@ and
@cv@ are not related by width subtyping: indeed, both vectors agree on
method names, but the types of the methods @getP1@ and @getP2@
differ. In @v@, the method @getP1@ has the type @IO@~@PrintablePoint@
whereas in @cv@ the same method has the type @IO@~@ColoredPoint@.
These different result types, @PrintablePoint@ and @ColoredPoint@,
are related by width subtyping.

The type of @cv@ is a \emph{deep} subtype of @v@. In \OOHaskell, we may
readily use functions (or methods) that exploit depth subtyping.  For
instance, we can define the following function for computing the norm
of a vector, and we can pass either vector @v@ or @cv@ to the
function.

\begin{code}
 norm v =
          do
              p1 <- v # getP1; p2 <- v  # getP2
              x1 <- p1 # getX; x2 <- p2 # getX
              return (abs (x1 - x2))
\end{code}
The above test code continues thus:

\begin{code}
                 -- ... continued ...
                 putStrLn "Length of v"
                 norm v >>= Prelude.print
                 putStrLn "Length of colored cv"
                 norm cv >>= Prelude.print
\end{code}

The method invocation operations within @norm@ remain well-typed no matter
which vector, @v@ or @cv@, we pass to that function. The
typing of @#@ is indeed compatible with both width and depth subtyping, and,
in fact, their combination. Thus, the object type $S$ is a subtype of
$T$ if the record type $S$ has all the fields of $T$ whose types are
not necessarily the same but related by subtyping in turn. Here we
assume, for now, that subtyping on method types is defined in
accordance to conservative rules~\cite{CW85,AC96}. (In the following
formulation, without loss of generality, we assume that \OOHaskell\ 
method types are monadic function types.) If $A_1
\to \cdots \to A_n \to \mathit{IO}\ R$ is a method type from $T$,
then there must be a method type in $S$, with the same method name,
and with a type $A'_1 \to \cdots \to A'_n \to \mathit{IO}\ R'$ such
that the following relationships hold:
\begin{itemize}
\item $A_1$, \ldots, $A_n$ must be subtypes of $A'_1$, \ldots, $A'_n$. \hfill (contra-variance)
\item $R'$ must be a subtype of $R$. \hfill (co-variance)
\end{itemize}
The above vector example exercises the co-variance of the result type
for the getters @getP1@ and @getP2@. 

We never had to specifically assert that the types of two objects are
related by width or depth subtyping. This is because in each and every
case, the compiler checks the well-typedness of all method invocations
directly, so no separate subtyping rules are needed. We contrast this
with type systems like System $F_{\le}$, where the subsumption rules
are explicitly asserted. The only place where an \OOHaskell\
programmer has to make the choice of subtyping relationship explicit
is in explicit narrowing operations. The previously described
operation @narrow@ covers width subtyping; the \OOHaskell\ library
also includes an operation @deep'narrow@. For instance, we can place
@v@ and @cv@, into the same homogeneous list:
\begin{code}
	let vectors = [v, deep'narrow cv]
\end{code}

The operation @deep'narrow@ descends into records, prefixes method
arguments by narrowing, and postfixes method results by narrowing.
Deep narrowing is just another record operation driven by the
structure of method types. (We refer to the source distribution for
details.) Deep narrowing is not the only way of dealing explicitly
with depth subtyping in \OOHaskell. We may also adopt the union-type
technique as of Sec.~\ref{S:union}.


\subsection{Co-variant method arguments}

{\sloppypar

\noindent
The variance of argument types is the subject of a significant
controversy~\cite{covariance-conflict,SG04,catcall}. The
contra-variant rule for method arguments entails type
substitutability, i.e., it assures the type safety of method
invocation for \emph{all} programs. However, argument type
contra-variance is known to be potentially too conservative. It is
often argued that a co-variant argument type rule is more suitable for
modelling real-world problems. If a method with co-variant argument
types happens to receive objects of expected types, then co-variance
is safe~---~for that particular program.  The proponents of the
co-variant argument type rule argue that because of the idiomatic
advantages of the rule we should admit it for those programs where it
is safe. It is the job of the compiler to warn the user when the
co-variant rule is used unsafely.  Alas, in the case of Eiffel~---~the
most established language with co-variance~---~the situation is the
following: ``No compiler currently available fully implements these
checks and behaviour in those cases ranges from run-time type errors
to system crashes.''~\cite{EiffelFAQ}.

}

In this section we demonstrate the restrictiveness of contra-variance
for method-argument types and show that \OOHaskell's subtyping
naturally supports type-safe co-variance. The faithful implementation
of the archetypal example from the Eiffel FAQ~\cite{EiffelFAQ} is
contained in the accompanying source code.

Continuing with the vector example from the previous section, we
extend @vector@ with a method, @moveO@ for moving the origin of the
vector. The method receives the new origin as a point object.

\begin{code}
 vector1 (p1::p) (p2::p) self =
   do
      super <- vector p1 p2 self
      returnIO
         $  moveO .=. (\pa -> do
                                 p1 <- self # getP1
                                 x  <- pa # getX
                                 p1 # moveX $ x)
        .*. super
\end{code}

As in the previous section, we construct the @vector1@ of plain
printable points @v1@ and the @vector1@ of colored points @cv1@. If we
intend @cv1@ to be substitutable for @v1@ in all circumstances (by the
virtue of depth subtyping), we must follow the contra-variance rule,
which requires the argument @pa@ of @moveO@ be either a plain
printable point (or an instance of its super-type). That requirement is
responsible for the longer-than-expected implementation of
@moveO@. Furthermore, the super-typing requirement on @pa@ precludes
@moveO@'s changing the color of the origin point, for the vector of
colored points. That degrades the expressiveness.

To illustrate the subtyping of the vectors, we define the function
that moves the origin of its vector argument to @0@: 
\begin{code}
 move_origin_to_0 varg = 
    do
       zero <- mfix (printable_point 0)
       varg # moveO $ zero
\end{code}
%
We may indeed apply that function to either @v1@ or @cv1@. The
function is polymorphic and can take any @vector1@ of plain points and
and its subtypes. The type of @cv1@ is truly a \emph{deep} subtype of the
type of @v1@. (Again, \OOHaskell\ does not require us to assert the
relevant subtype relationship in any way.)

We now turn to co-variant method argument types and so experiment with
yet another class of vectors. We also construct two instances of
@vector2@.

\begin{code}
 vector2 (p1::p) (p2::p) self =
   do
      p1r <- newIORef p1
      p2r <- newIORef p2
      returnIO $
           setO .=. writeIORef p1r
       -- ... other fields as in vector ...
\end{code}

\begin{code}
 testVector = do
                 -- ... test case as before ...
                 v2  <- mfix (vectors p1 p2)   -- vector of printable points
                 cv2 <- mfix (vectors cp1 cp2) -- vector of colored points
\end{code}
Like @vector1@, @vector2@ provides for setting the origin point; cf.\
the method @setO@. However, @vector2@ does that in a direct and simple
way; also, only @vector2@ permits changing the color of the origin
point, in a vector of colored points. Although the method @setO@ is
more convenient and powerful than the method @moveO@, the method
@setO@ has co-variant argument types~---~across printable-point and
colored-point vectors. For a vector of colored points, @cv2@, the
argument type of @setO@ must be a colored point too, i.e., the same
type as @p1r@~---~otherwise, the mutation @writeIORef@ cannot be
typed.

Hence, the type of @cv2@ cannot be a subtype of the type of @v2@
(because @setO@ breaks the contra-variant argument type rule). An OO
system that enforces the contra-variant rule will not allow us to
write functions that can take both @v2@ and @cv2@. For example, 
we may want to devise the following function:

\begin{code}
align_origins va vb = 
    do
       pa <- va # getP1
       vb # setO $ pa
\end{code}
It is always safe to apply @align_origins@ to two @vector2@s of the
same type. \OOHaskell\ does let us pass either two @vector2@s of
printable points (such as @v2@) to two @vector2@s of colored points
(such as @cv2@), and so vector types \emph{can be}
substitutable~---~despite a co-variant argument type of @set0@.

Substitutability is properly restricted for this function:

\begin{code}
 set_origin_to_0 varg = 
    do
       zero <- mfix (printable_point 0)
       varg # setO $ zero
\end{code}
%
We apply the function to @v2@, but if we try to apply it to @cv2@ we
get the type error message about a missing method @getColor@ (which
distinguishes colored points from plain printable points). Likewise,
we get an error if we attempt to place both @v2@ and @cv2@ in a
homogeneous list like this:

\begin{code}
	let vectors = [v2, deep'narrow cv2]
\end{code}
In this case, we can narrow both vectors to the type of @vector@
though, so that the offending method @setO@ will be projected out and
becomes private.

\OOHaskell\ typechecks actual operations on objects; therefore,
\OOHaskell\ permits methods with co-variant argument types in situations
where they are used safely. The type checker will flag any unsafe use
and force the programmer to remove the offending method. Permitting
safe uses of methods with co-variant argument types required no
programming on our part. We get this behaviour for free.


\subsection{Anti-patterns for subtyping}
\label{S:anti-patterns}

We have seen several approaches to the construction of a
subtype-polymorphic collection, as needed for the `scribble' loop in
the running shapes example. In the section on non-\OOHaskell\
encodings, Sec.~\ref{S:poor}, we had discussed two additional options:
\begin{itemize}
\item The use of \HList's heterogeneous lists.
\item The use of ``$\exists$'' to make the list element type opaque.
\end{itemize}
Albeit one might have expected these options to be of use, they turned
out to be problematic for OO programming with non-extensible Haskell
records.  In the combination with \OOHaskell\ (and its extensible
records), these two options are even less attractive.

In first approach, we construct the scribble list as:

\begin{code}
 let scribble = s1 `HCons` (s2 `HCons` HNil)
\end{code}
and use @hMap_@, Sec.~\ref{S:hetero}, to iterate over the list:

\begin{Verbatim}[fontsize=\small,commandchars=\\\{\}]
 hMapM_ (\undefined::FunOnShape) scribble
\end{Verbatim}
where there must be an instance of type class @Apply@ for @FunOnShape@,
e.g.:

\begin{code}
instance ( HasField (Proxy Draw) r (IO ())
         , HasField (Proxy RMoveTo) r (Int -> Int -> IO ())
         )
      => Apply FunOnShape r (IO ())
  where
    apply _ x = do
                   x # draw
                   (x # rMoveTo) 100 100
                   x # draw
\end{code}
Haskell's type class system requires us to provide proper bounds for
the instance, hence the list of the \emph{method-access constraints}
(for ``@#@'', i.e., @HasField@) above.  The form of these constraints
strongly resembles the method types listed in the shape interface
type, Sec.~\ref{S:narrow}. One may wonder whether we can
somehow use the full type synonym @Shape@, in order to constrain the
instance.  This is not possible in Haskell because constraints are not
first-class citizens in Haskell; we cannot compute them from types or
type proxies~---~unless we were willing to rely on heavy encoding or
advanced syntactic sugar. So we are doomed to manually infer and
explicitly list such method-access constraints for each such piece of
polymorphic code.

The existential quantification approach falls short for essentially
the same reason. Assuming a suitable existential envelope and
following Sec.~\ref{S:exists}, we can build @scribble@ as

\begin{code}
 let scribble = [ HideShape s1, HideShape s2 ]
\end{code}
The declaration of the existential type depends on the function that
we want to apply to the opaque data. When iterating over the list, via
@mapM_@, we only need to unwrap the @HideShape@ constructor prior to
method invocations:

\begin{code}
 mapM_ ( \(WrapShape shape) -> do
             shape # draw
             (shape # rMoveTo) 100 100
             shape # draw )
         scribble
\end{code}
These operations have to be anticipated in the type bound for the
envelope:

\begin{Verbatim}[fontsize=\small,commandchars=\\\{\}]
 data OpaqueShape =
  forall x. ( HasField (Proxy Draw) x (IO ())
            , HasField (Proxy RMoveTo) x (Int -> Int -> IO ())
            ) => HideShape x
\end{Verbatim}

This approach evidently matches the \HList-based technique in
terms of encoding efforts. In both cases, we need to identify
type class constraints that correspond to the (potentially)
polymorphic method invocations. This is impractical. Not even
mainstream OO languages with no advanced type inference, require this
sort of type information from the programmer.

Existential quantification can also be used for object encoding, e.g.,
for wrapping up @self@. That lets us, for example, easily implement
self-returning methods without resorting to infinite types. Such use of
existential quantification is not practical in \OOHaskell\ for the
same reason: it requires us to \emph{exhaustively} enumerate all
type classes an object and any of its types are or will be the
instances of.


\section{Discussion}
\label{S:discuss}

We will first discuss usability issues of the current \OOHaskell\
library, further constrained by current Haskell implementations. We
will then summarise related work on functional object-oriented
programming in Haskell and elsewhere. Finally, we will list topics for
future work~---~other than just improving usability of \OOHaskell.


\subsection{Usability issues}


\subsubsection{Usability of inferred types}

So far, we have not shown any type inferred by Haskell for our
objects. One may wonder how readable and comprehensible they are, if
they can be used as means of program understanding, and if a Haskell
language extension is needed to improve the presentation of the
inferred types. In upshot, the inferred types are reasonable for
simple OO programming examples, but there is a fuzzy borderline beyond
which the volume and the idiosyncrasies of inferred types injure
their usefulness. This concern suggests an important topic for future
work.

Let us see the inferred type of the colored point introduced in
Sec.~\ref{S:single-inheritance:extension}:
 
\begin{code}
 ghci6.4> :t mfix $ colored_point (1::Int) "red"
 mfix $ colored_point (1::Int) "red" ::
        IO (Record 
            (HCons (Proxy GetColor, IO String)
             (HCons (Proxy VarX, IORef Int)
              (HCons (Proxy GetX, IO Int)
               (HCons (Proxy MoveX, Int -> IO ())
                (HCons (Proxy Print, IO ())
                 HNil))))))
\end{code} 
We think that this type is quite readable, even though it reveals the
underlying representation of records (as a heterogeneous list of
label-value pairs), and gives away the proxy-based model for
labels.  We may hope for a future Haskell implementation whose
customisable `pretty printer' for types would present the result of
type inference perhaps as follows:

\begin{code}
 ghci> :t mfix $ colored_point (1::Int) "red"
 mfix $ colored_point (1::Int) "red" ::
        IO ( Record (
                  GetColor :=: IO String
              :*: VarX     :=: IORef Int
              :*: GetX     :=: IO Int
              :*: MoveX    :=: (Int -> IO ())
              :*: Print    :=: IO ()
              :*: HNil ))
\end{code}

The above example dealt with monomorphic objects. Let us also see the
inferred type of a polymorphic object generator, with `open recursion
left open'. Here is the (pretty-printed) type of the object generator
for colored points:

\begin{code}
 ghci> :t colored_point
 ( Num a
 , HasField (Proxy GetX) r (IO a1)
 , Show a1
 ) => a
   -> String
   -> r
   -> IO ( Record (
             GetColor :=: IO String
         :*: VarX     :=: IORef a
         :*: GetX     :=: IO a
         :*: MoveX    :=: (a -> IO ())
         :*: Print    :=: IO ()
         :*: HNil ))
\end{code}
The inferred type lists all the fields of an object, both new and
inherited. Assumptions about @self@ are expressed as constraints on
the type variable @r@. The object generator refers to @getX@ (through
@self@), which entails a constraint of the form
@HasField@~@(Proxy@~@GetX)@~@r@~@(IO@~@a1)@.  The coordinate type for
the point is polymorphic; cf.\ @a@ for the initial value and @a1@ for
the value retrieved by @getX@. Since arithmetics is performed on the
coordinate value, this implies bounded polymorphism: only @Num@-ber
types are permitted. We cannot yet infer that @a@ and @a1@ must
eventually be the same since `the open recursion is still open'.

We must admit that we have assumed a \emph{relatively} eager instance
selection in the previous Haskell session. The Hugs implementation of Haskell
is (more than) eager enough. The recent versions of GHC have become
quite lazy. In a session with contemporary GHC (6.4), the
inferred type would comprise the following additional constraints,
which all deal with the uniqueness of label sets as they are
encountered during record extension:

\begin{Verbatim}[fontsize=\small,commandchars=\\\{\}]
 HRLabelSet (HCons (Proxy MoveX, a -> IO ())
            (HCons (Proxy Print, IO ()) HNil)),
 \cmt{likewise for} MoveX, Print, GetX
 \cmt{likewise for} MoveX, Print, GetX, VarX
 \cmt{likewise for} MoveX, Print, GetX, VarX, GetColor
\end{Verbatim}
Inspection of the @HRLabelSet@ instances shows that these constraints
are all satisfied, no matter how the type variable @a@ is
instantiated. No ingenuity is required. A simple form of strictness
analysis were sufficient. Alas, GHC is consistently lazy in resolving
even such constraints. Modulo @HRLabelSet@ constraints, the inferred
type seems quite reasonable, explicitly listing all
relevant labels and types of the record components.


\subsubsection{Usability of type errors}

Due to \OOHaskell's extensive use of type-class-based programming,
there is a risk that type errors may become too complex. We will look
at some examples. The results clearly provide incentive for future
work on the subject of type errors.

Let us first attempt to instantiate an abstract class, e.g.,
@abstract_point@ from Sec.~\ref{S:constrain}. That object generator
defined the @print@ method, which invoked @getX@ on @self@. The latter
is left to be defined in concrete subclasses. If we take the fixpoint of
such an `incomplete' object generator, Haskell's type checker (here:
GHC 6.4) gives the following error message:

\begin{code}
 ghci> let x = mfix (abstract_point 7)
 No instance for (HasField (Proxy GetX) HNil (IO a))
   arising from use of `abstract_point' at <interactive>:1:14-27
 Probable fix:
   add an instance declaration for (HasField (Proxy GetX) HNil (IO a1))
 In the first argument of `mfix', namely `(abstract_point 7)'
 In the definition of `x': x = mfix (abstract_point 7)
\end{code}
We think that the error message is concise and to the point. The
message succinctly lists just the missing field (The suggested
`probable fix' is not really helpful here). In our next scenario, we
use a version of @abstract_point@ that comprises an instantiation test
by constraining @self@ through @narrow@, as discussed in
Sec.~\ref{S:constrain}:

\begin{code}
 abstract_point (x_init::a) self =
   do
      ... as before ...
 where
  _ = narrow self :: Record (  GetX  :=: IO a
                           :*: MoveX :=: (a -> IO ())
                           :*: HNil )
\end{code}
When we now take the fixpoint again, we get a more complex error message:

\begin{code}
 ghci> let x = mfix (abstract_point 7)
 No instance for (HExtract HNil (Proxy GetX) (IO a),
                  HExtract HNil (Proxy MoveX) (a -> IO ()),
                  HasField (Proxy GetX) HNil (IO a1))
   arising from use of `abstract_point' at <interactive>:1:14-27
 Probable fix: ...
 In the first argument of `mfix', namely `(abstract_point 7)'
 In the definition of `x': x = mfix (abstract_point 7)
\end{code}
Compared to the earlier error message, there are two additional
unsatisfied @HExtract@ constraints. Two out of the three constraints
refer to @GetX@, and they complain about the same problem: a missing
method implementation for @getX@. The constraint regarding @MoveX@
deals with a pure virtual method that is not used in the object
generator. The kinds and numbers of error messages for @GetX@ and
@MoveX@ may lead to confusion; internals of \OOHaskell\ end up at the
surface.

In order to improve on such problems, the Haskell type system and its
error-handling part would need to be opened up to allow for
problem-specific error messages. We would like to refine Haskell's
type checker so that type error messages directly refer to the
involved OO concepts.

Let us consider yet another scenario. We turn to self-returning
methods, as we discussed them in Sec.~\ref{S:self}. In the following
flawed \OOHaskell\ program, we attempt to return @self@ right away:

\begin{code}
 self_returning_point (x_init::a) self =
   do
      super <- printable_point x_init self
      returnIO
          $  me .=. self -- assumes iso-recursive types
         .*. super
\end{code}
%
The problem will go unnoticed until we try to @mfix@ the
generator, at which point we get a type error:

{\footnotesize

\begin{code}
    Occurs check: cannot construct the infinite type:
      a
      =
      Record (HCons (Proxy Me, a)
              (HCons (Proxy MutableX, IORef a1)
               (HCons (Proxy GetX, IO a1)
                (HCons (Proxy MoveX, a1 -> IO ())
                 (HCons (Proxy Print, IO ()) HNil)))))
      Expected type: a -> IO a
      Inferred type: a -> IO (Record (HCons (Proxy Me, a)
                      (HCons (Proxy MutableX, IORef a1)
                       (HCons (Proxy GetX, IO a1)
                        (HCons (Proxy MoveX, a1 -> IO ())
                         (HCons (Proxy Print, IO ()) HNil))))))
    In the application `self_returning_point 7'
    In the first argument of `mfix', namely `(self_returning_point 7)'
\end{code}

}

This error message is rather complex compared to the simple object
types that are involved. Although the actual problem is correctly
described, the programmer receives no help in locating the offending
code, @me .=. self@. The volume of the error message is the
consequence of our use of structural types. One may think that adding
some type synonyms and using them in type signatures should radically
improve the situation. It is true that contemporary Haskell type
checkers keep track of type synonyms. However, an erroneous
subexpression may just not be sufficiently annotated or constrained by
its context. Also, the mere coding of type synonyms is very
inconvenient. This situation suggests that a future Haskell type
checker could go two steps further. Our first proposal is to allow for
the inference of type synonyms; think of:

\begin{code}
 foo x y z = ...  -- complex expression on structural object types
 type Foo  = typeOf foo            -- capture the type in an alias
\end{code}

\noindent
(Here, @typeOf@ is an envisaged extension.)  Our second proposal is to
use type synonyms aggressively for the simplification of inferred
types or type portions of error messages. This is a challenging
subject given Haskell's forms of polymorphism.

The verbosity of \OOHaskell\ error messages may occasionally compare
to error messages in C++ template instantiation, which can be
immensely verbose, spanning several dozens of packed lines, and yet
boost and similar C++ libraries, which extensively use templates, are
gaining momentum. In general, the clarity of error messages is
undoubtedly an area that needs more research, and such research is
being carried out by Sulzmann and others~\cite{SSW04}, which
\OOHaskell\ programmers and Haskell compiler writers may take
advantage of.

The ultimate conclusion of our discussion of inferred types and type
errors is that such type information needs to be presented to the
programmer in an abbreviated and OO-aware fashion. This proposal is
based on the observation of OCaml's development. Although objects
types shown by OCaml are quite concise, that has not always been the
case. In the {ML-ART} system, the predecessor of OCaml with no
syntactic sugar~\cite{ML-ART}, the printed inferred types were not
unlike the \OOHaskell\ types we have seen in this section.
\begin{quote}
``Objects have anonymous, long, and often recursive
types that describe all methods that the object can receive. Thus, we
usually do not show the inferred types of programs in order to
emphasise object and inheritance encoding rather than typechecking
details. This is quite in a spirit of ML where type information is
optional and is mainly used for documentation or in module
interfaces. Except when trying top-level examples, or debugging, the
user does not often wish to see the inferred types of his programs in
a batch compiler.''
\end{quote}


\subsubsection{Efficiency of object encoding}
\label{S:efficiency}

Our representation of objects and their types is \emph{deliberately}
straightforward: polymorphic, extensible records of closures.  This
approach has strong similarities with prototype-based systems (such as
Self~\cite{Self}) in that mutable fields and method `pointers' are
contained in one record. A more efficient representation based on
separate method and field tables (as in C++ and Java) is possible, in
principle. Although our current encoding is certainly not optimal, it
is conceptually clearer. This encoding is used in such languages as
Perl, Python, Lua~---~and is often the first one chosen when adding OO
to an existing language. 

The efficiency of the current \OOHaskell\ encoding is also problematic
for reasons other than separation of fields and methods. For example,
although record extension is constant (run-)time, the field/method
lookup is linear search. Clearly, a more efficient encoding is
possible: one representation of the labels in the \HList\ paper
permits a total order among the labels types, which in turn, permits
construction of efficient search trees. We may also impose an order on
the components per record type, complete with subtype-polymorphic
record extension only to the right, so that labels can be mapped to
array indexes.

In the present paper, we chose conceptual clarity over such
optimisations.  Furthermore, a non-trivial case study is needed to
drive optimisations. Mere improvements in object encoding may be
insufficient however. The compilation time of \OOHaskell\ programs and
their runtime efficiency is challenged by the number of heavily nested
dictionaries that are implied by our systematic type-class-based
approach. It is quite likely that a scalable \HList/\OOHaskell\ style
of programming will require compiler optimisations that make
type-class-based programming more efficient~---~in general.


\subsection{Related work}
\label{S:related}

Throughout the paper we referenced related work whenever specific
technical aspects suggested to do so. We will complete the picture by
a broader discussion. There are three overall dimensions of related
work: foundations of object encoding (cf.\ Sec.~\ref{S:MLART}),
Haskell extensions for OO (cf.\ Sec.~\ref{S:extension}), and OO
encoding in Haskell (cf.\ Sec.~\ref{S:encoding}).

The literature on object encoding is quite extensive. \OOHaskell\
takes advantage of seminal work such
as~\cite{CW85,AC96,Ohori95,PT94,BM92}. Most often, typed object
encodings are based on polymorphic lambda calculi with subtyping,
while there are also object calculi that start, more directly, from
objects or records. Due to this overwhelming variety, we narrow down
the discussion. We identify {ML-ART}~\cite{ML-ART} by R{\'e}my et al.\
(see also~\cite{RV97}) as the closest to \OOHaskell~---~in motivation
and spirit, but not in the technical approach. Hence,
Sec.~\ref{S:MLART} is entirely focused on ML-ART, without further
discussion of less similar object encodings. The distinguishing
characteristic of \OOHaskell\ is the use of type-class-bounded
polymorphism.


\subsubsection{The ML-ART object encoding}
\label{S:MLART}

Both ML-ART and \OOHaskell\ identify a small set of language features
that make functional object-oriented programming possible. In both
projects, the aim was to be able to implement objects~---~as a
\emph{library} feature. Therefore, several OO styles can be
implemented, for different classes of users and classes of
problems. One does not need to learn any new language and can discover
OO programming progressively. Both {ML-ART} and \OOHaskell\ base their
object systems on polymorphic extensible records. Both \OOHaskell\ and
{ML-ART} deal with mutable objects (\OOHaskell\ currently neglects
functional objects since they are much less commonly used in
practise). Both \OOHaskell\ and {ML-ART} aim at preserving type
inference.

ML-ART adds several extensions to ML to implement objects: records
with polymorphic access and extension, projective records, recursive
types, implicit existential and universal types. As the ML-ART
paper~\cite{ML-ART} reports, none of the extensions are new, but their
combination is original and ``provides just enough power to program
objects in a flexible and elegant way''.

We make the same claim for \OOHaskell, but using a quite different set
of features. What fundamentally sets us apart from {ML-ART} is the
different source language: Haskell. In Haskell, we can implement
polymorphic extensible records natively rather than via an
extension. We use type-class-based programming to this end.\footnote{
The fact that such records are realisable in Haskell at all has been
unknown, until the \HList\ paper, which we published in 2004. The
assumed lack of extensible records in Haskell was selected as prime
topic for discussion at the Haskell 2003 workshop~\cite{HW03}.} We
avoid row variables and their related complexities. Our records permit
introspection and thus let us \emph{implement} various type-safe cast
operations appealing to different subtyping relationships.  For
instance, unlike {ML-ART}, \OOHaskell\ can compute the most common type
of two record types without requiring type annotations. Quoting
from the ML-ART paper:

\begin{quote}
``The same message print can be sent to points and colored
points. However, both of them have incompatible types and can never be
stored in the same list. Some languages with subtyping allow this
set-up. They would take the common interface of all objects that are
mixed in the list as the interface of any single object of the list.''
\end{quote}

Unlike ML-ART, we do \emph{not} rely on existential or implicitly
universal types, nor recursive types. We use value recursion
instead. That representation, a record of recursive closures,
abstracts the internal state of the object~---~its value as well its
type. Haskell helps us overcome what ML-ART calls ``severe
difficulties'' with value recursion. In ML, the difficulties are
serious enough to abandon the value recursion, despite its attractive
features in supporting implicit subtyping, in favour of more complex
object encodings requiring extensions of the type system. The subtle
problem of value recursion is responsible for complicated and
elaborate rules of various mainstream OO languages that prescribe what
an object constructor may or may not do.  The ML-ART paper
mentions an unpublished attempt (by Pierce) to take advantage of the
facts that fixpoints in a call-by-name language are always safe and
that call-by-name can be emulated in a call-by-value language with the
help of extra abstraction (thunks). However, in that attempted
implementation the whole message table had to be rebuilt every time an
object sends a message to self and so that approach was not pursued
further. Our simple scheme of Sec.~\ref{S:mfix-safety} seems to answer
the ML-ART challenge~---~``to provide a clean and efficient solution that
permits restricted form of recursion on non-functional values.''

ML-ART uses a separate method table, whereas \OOHaskell\ uses a single
record for both mutable fields and method `pointers'.  The ML-ART
encoding is more efficient than that of \OOHaskell. All instances of
an object (class) literally share the same method table. ML-ART (and
OCaml) is also more efficient simply because more elements of the
object encoding are natively implemented. By contrast, \OOHaskell's
type system is programmed through type-class-based programming. As a
result, \OOHaskell is definitely less fit for \emph{practical} OO software
development than ML-ART (or rather OCaml).


\subsubsection{Haskell language extensions}
\label{S:extension}

There were attempts to bring OO to Haskell by a language extension. An
early attempt is Haskell++~\cite{HS95} by Hughes and Sparud. The
authors motivated their extension by the perception that Haskell lacks
the form of incremental reuse that is offered by inheritance in
object-oriented languages. Our approach uses common extensions of the
Hindley/Milner type system to provide the key OO notions.  So in a
way, Haskell's fitness for OO programming just had to be discovered,
which is the contribution of this paper.

O`Haskell~\cite{Nordlander98,Nordlander02} is a comprehensive OO
variation on Haskell designed by Nordlander. O`Haskell extends Haskell
with reactive objects and subtyping. The subtyping part is a
substantial extension. The reactive object part combines stateful
objects and concurrent execution, again a major extension. Our
development shows that no extension of Haskell is necessary for
stateful objects, and the details of the object system can be
programmed in Haskell.

Another relevant Haskell variation is Mondrian. In the original paper
on the design and implementation of Mondrian~\cite{MC97}, Meijer and
Claessen write: ``The design of a type system that deals with
subtyping, higher-order functions, and objects is a formidable
challenge ...''. Rather than designing a very complicated language,
the overall principle underlying Mondrian was to obtain a simple
Haskell dialect with an object-oriented flavour. To this end,
algebraic datatypes and type classes were combined into a simple
object-oriented type system with no real subtyping, with completely
co-variant type-checking. In Mondrian, runtime errors of the kind
``message not understood'' are considered a problem akin to partial
functions with non-exhaustive case discriminations. \OOHaskell\ raises
the bar by providing proper subtyping (``all message will be
understood'') and other OO concepts in Haskell without extending the
Haskell type system.


\subsubsection{Object encodings for Haskell}
\label{S:encoding}

This paper may claim to provide the most authoritative analysis of
possible object encodings in Haskell; cf.\ Sec.~\ref{S:poor}. Previous
published work on this subject has not addressed general (functional)
object-oriented programming, but it has focused instead on the import
of foreign libraries or components into
Haskell~\cite{FLMPJ99,SPJ01,PC04}. The latter problem domain makes
important simplifying assumptions:

\begin{itemize}\noskip
\item Object state does not reside in Haskell data.
\item There are only (opaque) object ids referring to the foreign site.
\item State is solely accessed through methods (``properties'').
\item Haskell methods are (often generated) stubs for foreign code.
\item As a result, such OO styles just deal with interfaces.
\item No actual (sub)classes are written by the programmer.
\end{itemize}

In this restricted context, one approach is to use phantom types for
recording inheritance relationships~\cite{FLMPJ99}. Each interface is
represented by an (empty) datatype with a type parameter for
extension. After due consideration, it turns out that this approach is
a restricted version of what Burton called ``type extension through
polymorphism'': even records can be made extensible through the
provision of a polymorphic dummy field~\cite{Burton90}. Once we do not
maintain Haskell data for objects, there is no need to maintain a
record type, but the extension point is a left over, and it becomes a
phantom. We have ``re-generalised'' the phantom approach in
Sec.~\ref{S:burton}.

{\sloppypar

Another approach is to set up a Haskell type class to represent the
subtyping relationship among interfaces~\cite{SPJ01,PC04} where each
interface is modelled as a dedicated (empty) Haskell type. We have
enhanced this approach by state in Sec.~\ref{S:objcomp}.

}

Based on our detailed analysis of both approaches, we submit that the
second approach seems to be slightly superior to the first one, while
both approaches are too cumbersome for actual functional OO
programming.

{\sloppypar

Not in peer-referred publications, but in Haskell coding practise,
some sorts of OO-like encodings are occasionally found. For instance,
it is relatively well understood that Haskell's type classes allow for
interface polymorphism or for abstract classes (type classes) vs.\
concrete classes (type class instances). As of writing, the published
Haskell reference solution for the shapes example,
\url{http://www.angelfire.com/tx4/cus/shapes/}, is a
simple-to-understand encoding that does not attempt to maximise reuse
among data declarations and accessors. The encoding is specialised to
the specific problem; the approach may fail to scale.  The encoding
also uses existentials for handling subtype-polymorphic collections,
which is an inherently problematic choice, as we have shown in
Sec.~\ref{S:anti-patterns}.

}


\subsection{More future work}

We have focused on mutable objects so far; studying functional objects
appears to be a natural continuation of this work, even though
functional objects are of much less practical relevance. \omitnow{The
  use of two tables (one for methods, another for fields, as discussed
  in Sec.~\ref{S:efficiency}) not only helps with efficiency, but also
  simplifies the provision of functional objects. The source
  distribution comprises some bits that hint at the possible encoding
  of functional objects.}

The notion of object construction as a multi-stage computation (cf.\
Sec.~\ref{S:mfix-safety}) merits further exploration (as well as the
clarification of the relationship with environment
classifiers~\cite{env-classifiers}).

\OOHaskell\ should be elaborated to cover general forms of reflective
programming and, on the top of that, general forms of aspect-oriented
programming. A simple form of reflection is already provided in terms
of the type-level encoding of records. We can iterate over records and
their components in a generic fashion. Further effort is needed to
cover more advanced forms of reflection such as the iteration over the
object pool, or the modification of object generators.

Another promising elaboration of \OOHaskell\ would be its use for
the reusable representation of design-pattern solutions.


\section{Concluding remarks}
\label{S:concl}

The present paper addresses the intellectual challenge of seeing if
the conventional OO idioms can at all be implemented in Haskell (short
of writing a compiler for an OO language in Haskell). Peyton Jones and
Wadler's paper on imperative programming in
Haskell~\cite{peytonjoneswadler-popl93} epitomises such an
intellectual tradition for the imperative paradigm. The same kind of
intellectual challenge, `paradigm assimilation', is addressed by
FC++~\cite{fcpp-jfp}, which implements in C++ the quintessential
Haskell features: type inference, higher-order functions,
non-strictness. The present paper, conversely, faithfully (i.e., in a
similar syntax and without global program transformation) realises a
principal C++ trait~---~OO programming.  According to Peyton Jones,
Haskell is ``the world's finest imperative programming
language''~\cite{spj00}. We submit that Haskell is also a
\emph{bleeding-edge OO programming language}, while we readily restrict this
claim to mere OO language-design capability; much more work would be
needed to enable scalable OO software development with Haskell.

We have discovered an object system for Haskell that supports stateful
objects, inheritance and subtype polymorphism. We have implemented OO
as a Haskell library, \OOHaskell, based on the polymorphic, extensible
records with introspection and subtyping provided by the \HList\
library~\cite{HLIST-HW04}. Haskell programmers can use OO idioms if it
suits the problem at hand. We have demonstrated that \OOHaskell\
programs are very close to the textbook OO code, which is normally
presented in mainstream OO languages. \OOHaskell's deviations are
appreciated. The \OOHaskell\ library offers a comparatively rich
combination of OO idioms. Most notably, we have implemented
parameterised classes, constructor methods, abstract classes, pure
virtual methods, single inheritance, multiple inheritance, object
composition, structural types, and nominal types.  The choice of
Haskell as a base language has allowed us to deliver extensive type
inference, first-class classes, implicit polymorphism of classes, and
more generally: programmable OO type systems. Starting from the
existing \OOHaskell\ library and the corresponding sample suite, one
can explore OO language design, without the need to write a compiler.

The present paper settles the question that hitherto has been
open. The conventional OO idioms in their full generality
\emph{are} expressible in current Haskell without any new
extensions. It turns out, Haskell~98 plus multi-parameter type classes
with functional dependencies are sufficient. This combination is
well-formalised and reasonably understood~\cite{SS05}. Even
overlapping instances are not essential (yet using them permits a more
convenient representation of labels, and a more concise implementation
of some type-level functionality). The fact that we found a quite
unexpected (and unintended) use of the existing Haskell features is
reminiscent of the accidental discovery of C++ template
meta-programming. The latter is no longer considered an exotic
accident or a type hack~---~rather, a real feature of the
language~\cite{DSL-in-three-lang}, used in the Standard Template
Library and described in popular C++ books, e.g., \cite{Alexandrescu01}.

Haskell has let us move beyond the mere curiosity of implementing OO
idioms to the point of making contributions to open and controversial
OO problems. Haskell has let us concisely specify and enforce the
restrictions on the behaviour of object constructors (preventing the
constructor access not-yet-fully constructed objects). The object
encoding with recursive records \emph{can} be made safe. Also, we were
able to effortlessly implement fine-grain notions of width and depth
subtyping, with respect to particular object operations, and thus
safely permit methods with co-variant argument subtyping. Not only
\OOHaskell\ is able to automatically compute the least general
interface of a heterogeneous collection of objects (through 
semi-implicit upcasts) and make the collection homogeneous, but it
provides the means for safe downcasts. Moreover, downcasts that cannot
possibly succeed are flagged as type errors. These are capabilities
that go beyond state-of-the-art functional object-oriented programming
with OCaml.

{\sloppypar

Just as C++ has become the laboratory for generative programming
\cite{DSL-in-three-lang} and lead to such applications as
FC++~\cite{fcpp-jfp} and Boost (\url{http://www.boost.org/}), we
contend that (OO)Haskell would fit as \emph{the} laboratory for
advanced and typed OO language design. All our experiments have shown
that (OO)Haskell\ indeed supports a good measure of
experimentation~---~all without changing the type system and the
compiler.

}


{\small 

\subsubsection*{Acknowledgements}
 
We thank Keean Schupke for his major contributions to the \HList\ and
OOHaskell\ libraries. We thank Chung-chieh Shan for very helpful
discussions.  We also gratefully acknowledge feedback from Robin
Green, Bryn Keller, Chris Rathman and several other participants in
mailing list or email discussions. The second author presented this
work at an earlier stage at the WG2.8 meeting (Functional Programming)
in November 2004 at West Point. We are grateful for feedback received
at this meeting.

}


{

\bibliographystyle{thisIsNotJFP}
\bibliography{paper}

\begin{thebibliography}{}

\bibitem[\protect\citename{Abadi \& Cardelli, }1996]{AC96}
Abadi, M., \& Cardelli, L. (1996).
\newblock {\em {A Theory of Objects}}.
\newblock Monographs in Computer Science.
\newblock New York, NY: Springer-Verlag.

\bibitem[\protect\citename{Abadi {\em et~al.}\relax, }1989]{ACPP89}
Abadi, M., Cardelli, L., Pierce, B., \& Plotkin, G. 1989 (Jan.).
\newblock {Dynamic typing in a statically-typed language}.
\newblock {\em Pages  213--227 of:} {\em {16th ACM Conference on Principles of
  Programming Languages}}.

\bibitem[\protect\citename{Abadi {\em et~al.}\relax, }1991]{ACPP91}
Abadi, M., Cardelli, L., Pierce, B., \& Plotkin, G. (1991).
\newblock {Dynamic typing in a statically-typed language}.
\newblock {\em {TOPLAS}}, {\bf 13}(2), 237--268.

\bibitem[\protect\citename{Aho {\em et~al.}\relax, }1986]{ASU86}
Aho, A.V., Sethi, R., \& Ullman, J.D. (1986).
\newblock {\em Compilers. principles, techniques and tools}.
\newblock Addison-Wesley.

\bibitem[\protect\citename{Alexandrescu, }2001]{Alexandrescu01}
Alexandrescu, A. (2001).
\newblock {\em {Modern C++ Design}}.
\newblock Pearson Education.

\bibitem[\protect\citename{Bayley, }2005]{MonadReader3}
Bayley, A. 2005 (June).
\newblock {\em {Functional Programming vs.\ Object Oriented Programming}}.
\newblock Monad.Reader, \url{http://www.haskell.org/tmrwiki/FpVsOo}.

\bibitem[\protect\citename{Bruce \& Mitchell, }1992]{BM92}
Bruce, K.B., \& Mitchell, J.C. (1992).
\newblock {PER models of subtyping, recursive types and higher-order
  polymorphism}.
\newblock {\em Pages  316--327 of:} {\em {POPL 1992: Proc.\ of the 19th ACM
  SIGPLAN-SIGACT symposium on Principles of programming languages}}.
\newblock ACM Press.

\bibitem[\protect\citename{Bruce {\em et~al.}\relax, }2003]{PolyTOIL}
Bruce, K.B., Schuett, A., {van Gent}, R., \& Fiech, A. (2003).
\newblock {PolyTOIL: A type-safe polymorphic object-oriented language}.
\newblock {\em {TOPLAS}}, {\bf 25}(2), 225--290.

\bibitem[\protect\citename{Burton, }1990]{Burton90}
Burton, F.W. (1990).
\newblock {Type extension through polymorphism}.
\newblock {\em {TOPLAS}}, {\bf 12}(1), 135--138.

\bibitem[\protect\citename{Cardelli \& Wegner, }1985]{CW85}
Cardelli, L., \& Wegner, P. (1985).
\newblock {On Understanding Types, Data Abstraction, and Polymorphism}.
\newblock {\em {ACM Computing Surveys}}, {\bf 17}(4), 471--522.

\bibitem[\protect\citename{Castagna, }1995]{covariance-conflict}
Castagna, G. (1995).
\newblock Covariance and contravariance: Conflict without a cause.
\newblock {\em {TOPLAS}}, {\bf 17}(3), 431--447.

\bibitem[\protect\citename{Chen {\em et~al.}\relax, }1992]{CHO92}
Chen, K., Hudak, P., \& Odersky, M. (1992).
\newblock {Parametric type classes}.
\newblock {\em Pages  170--181 of:} {\em {Proceedings of the 1992 ACM
  Conference on LISP and Functional Programming}}.
\newblock ACM Press.

\bibitem[\protect\citename{{comp.lang.eiffel}, }2004]{EiffelFAQ}
{comp.lang.eiffel}. 2004 (17~Apr.).
\newblock {\em {Frequently Asked Questions (FAQ)}}.
\newblock \url{http://www.faqs.org/faqs/eiffel-faq/}.

\bibitem[\protect\citename{Cook, }1989]{CookThesis}
Cook, W.~R. (1989).
\newblock {\em {A Denotational Semantics of Inheritance}}.
\newblock Ph.D. thesis, Brown University.

\bibitem[\protect\citename{Cook {\em et~al.}\relax, }1990]{CHC90}
Cook, W.R., Hill, W., \& Canning, P.S. (1990).
\newblock {Inheritance is not subtyping}.
\newblock {\em Pages  125--135 of:} {\em {POPL '90: Proceedings of the 17th ACM
  SIGPLAN-SIGACT Symposium on Principles of Programming Languages}}.
\newblock New York, NY, USA: ACM Press.

\bibitem[\protect\citename{Czarnecki {\em et~al.}\relax,
  }2003]{DSL-in-three-lang}
Czarnecki, K., O'Donnell, J.T., Striegnitz, J., \& Taha, W. (2003).
\newblock {DSL} implementation in {MetaOCaml}, {Template} {Haskell}, and {C++}.
\newblock {\em Pages  51--72 of:} Lengauer, C., Batory, D.S., Consel, C., \&
  Odersky, M. (eds), {\em Domain-specific program generation}.
\newblock LNCS, vol. 3016.
\newblock Springer-Verlag.

\bibitem[\protect\citename{Duck {\em et~al.}\relax, }2004]{DPJSS04}
Duck, G.J., {Peyton Jones}, S.L., Stuckey, P.J., \& Sulzmann, M. (2004).
\newblock {Sound and Decidable Type Inference for Functional Dependencies}.
\newblock {\em Pages  49--63 of:} Schmidt, D.A. (ed), {\em {Proceedings, 13th
  European Symposium on Programming, ESOP 2004, Barcelona, Spain, March 29 -
  April 2, 2004}}.
\newblock LNCS, vol. 2986.
\newblock Springer-Verlag.

\bibitem[\protect\citename{Finne {\em et~al.}\relax, }1999]{FLMPJ99}
Finne, S., Leijen, D., Meijer, E., \& {Peyton Jones}, S.L. (1999).
\newblock {Calling hell from heaven and heaven from hell}.
\newblock {\em Pages  114--125 of:} {\em {ICFP '99: Proceedings of the fourth
  ACM SIGPLAN International Conference on Functional Programming}}.
\newblock New York, NY, USA: ACM Press.

\bibitem[\protect\citename{Gamma {\em et~al.}\relax, }1994]{GHJV94}
Gamma, E., Helm, R., Johnson, R., \& Vlissides, J. (1994).
\newblock {\em Design patterns: Elements of reusable object-oriented software}.
\newblock Addison-Wesley.

\bibitem[\protect\citename{Gaster \& Jones, }1996]{GJ96}
Gaster, B.R., \& Jones, M.P. 1996 (Nov.).
\newblock {\em {A Polymorphic Type System for Extensible Records and
  Variants}}.
\newblock Technical report NOTTCS-TR-96-3. University of Nottingham, Department
  of Computer Science.

\bibitem[\protect\citename{{H.~Nilsson}, }2003]{HW03}
{H.~Nilsson}. (2003).
\newblock {\em {The Future of Haskell discussion at the Haskell Workshop}}.
\newblock Message on the Haskell mailing list,
  \url{http://www.mail-archive.com/haskell\@haskell.org/msg13366.html}.

\bibitem[\protect\citename{Hallgren, }2001]{Hallgren01}
Hallgren, T. (2001).
\newblock {Fun with functional dependencies}.
\newblock  {\em {Joint Winter Meeting of the Departments of Science and
  Computer Engineering, Chalmers University of Technology and Goteborg
  University, Varberg, Sweden, Jan.\ 2001}}.
\newblock \url{http://www.cs.chalmers.se/~hallgren/Papers/wm01.html}.

\bibitem[\protect\citename{Howard {\em et~al.}\relax, }2003]{catcall}
Howard, M., Bezault, E., Meyer, B., Colnet, D., Stapf, E., Arnout, K., \&
  Keller, M. 2003 (27~Apr.).
\newblock {\em {Type-safe covariance: Competent compilers can catch all
  catcalls}}.
\newblock Work done as part of Eiffel language standardization by the TC39-TG4
  committee of ECMA; Draft; Available at
  \url{http://www.inf.ethz.ch/~meyer/ongoing/covariance/recast.pdf}.

\bibitem[\protect\citename{Hughes, }2002]{Hughes02}
Hughes, J. 2002 (23~Aug.).
\newblock {\em {Re: More suitable data structure needed}}.
\newblock Message on the Haskell mailing list,
  \url{http://www.haskell.org/pipermail/haskell/2002-August/010335.html}.

\bibitem[\protect\citename{Hughes \& Sparud, }1995]{HS95}
Hughes, J., \& Sparud, J. (1995).
\newblock {Haskell++: An Object-Oriented Extension of Haskell}.
\newblock  {\em {Proc.\ of Haskell Workshop, La Jolla, California}}.
\newblock {YALE Research Report DCS/RR-1075}.

\bibitem[\protect\citename{Jones, }1992]{MPJ92}
Jones, M.P. (1992).
\newblock {A theory of qualified types}.
\newblock {\em Pages  287--306 of:} {\em {Symposium proceedings on 4th European
  Symposium on Programming}}.
\newblock Springer-Verlag.

\bibitem[\protect\citename{Jones, }1995]{MPJ95}
Jones, M.P. (1995).
\newblock {Simplifying and improving qualified types}.
\newblock {\em Pages  160--169 of:} {\em {Proceedings of the 7th International
  Conference on Functional Programming Languages and Computer Architecture}}.
\newblock ACM Press.

\bibitem[\protect\citename{Jones, }2000]{MPJ00}
Jones, M.P. (2000).
\newblock {Type Classes with Functional Dependencies}.
\newblock {\em Pages  230--244 of:} {\em {Proceedings of the 9th European
  Symposium on Programming Languages and Systems}}.
\newblock Springer-Verlag.

\bibitem[\protect\citename{Kiselyov {\em et~al.}\relax, }2004]{HLIST-HW04}
Kiselyov, O., L{\"a}mmel, R., \& Schupke, K. (2004).
\newblock {Strongly typed heterogeneous collections}.
\newblock  {\em {ACM SIGPLAN Workshop on Haskell}}.
\newblock ACM Press.
\newblock See \url{http://www.cwi.nl/~ralf/HList/} for an extended technical
  report and for the source distribution.

\bibitem[\protect\citename{L{\"a}mmel \& {Peyton Jones}, }2003]{LPJ03}
L{\"a}mmel, R., \& {Peyton Jones}, S.L. (2003).
\newblock Scrap your boilerplate: a practical design pattern for generic
  programming.
\newblock {\em {ACM SIG{\-}PLAN Notices}}, {\bf 38}(3), 26--37.
\newblock Proc.\ of the ACM SIGPLAN Workshop TLDI~2003.

\bibitem[\protect\citename{Launchbury \& {Peyton Jones}, }1995]{LPJ95}
Launchbury, J., \& {Peyton Jones}, S.L. (1995).
\newblock State in {Haskell}.
\newblock {\em {Lisp and Symbolic Computation}}, {\bf 8}(4), 293--342.

\bibitem[\protect\citename{Leroy {\em et~al.\ }\relax, }2004]{OCaml}
Leroy, Xavier, {\em et~al.\ }\relax. 2004 (July).
\newblock {\em {The Objective Caml system, release 3.08, Documentation and
  user's manual}}.
\newblock \url{http://caml.inria.fr/ocaml/htmlman/index.html}.

\bibitem[\protect\citename{Liskov \& Wing, }1994]{LiskovW94}
Liskov, B., \& Wing, J.M. (1994).
\newblock {A Behavioral Notion of Subtyping}.
\newblock {\em {TOPLAS}}, {\bf 16}(6), 1811--1841.

\bibitem[\protect\citename{McBride, }2002]{Fake}
McBride, C. (2002).
\newblock {Faking It (Simulating Dependent Types in Haskell)}.
\newblock {\em {Journal of Functional Programming}}, {\bf 12}(4--5), 375--392.

\bibitem[\protect\citename{McNamara \& Smaragdakis, }2004]{fcpp-jfp}
McNamara, B., \& Smaragdakis, Y. (2004).
\newblock {Functional Programming with the {FC++} Library}.
\newblock {\em {Journal of Functional Programming}}, {\bf 14}(4), 429--472.

\bibitem[\protect\citename{Meijer \& Claessen, }1997]{MC97}
Meijer, E., \& Claessen, K. (1997).
\newblock {The Design and Implementation of Mondrian}.
\newblock  {\em {ACM SIGPLAN Haskell Workshop}}.
\newblock ACM Press.

\bibitem[\protect\citename{Neubauer {\em et~al.}\relax, }2001]{NTGS01}
Neubauer, M., Thiemann, P., Gasbichler, M., \& Sperber, M. (2001).
\newblock {A Functional Notation for Functional Dependencies}.
\newblock {\em Pages  101--120 of:} {\em {Proc.\ 2001 ACM SIGPLAN Haskell
  Workshop, Firenze, Italy, September 2001}}.

\bibitem[\protect\citename{Neubauer {\em et~al.}\relax, }2002]{NTGS02}
Neubauer, M., Thiemann, P., Gasbichler, M., \& Sperber, M. (2002).
\newblock {Functional logic overloading}.
\newblock {\em Pages  233--244 of:} {\em {Proceedings of the 29th ACM
  SIGPLAN-SIGACT Symposium on Principles of Programming Languages}}.
\newblock ACM Press.

\bibitem[\protect\citename{Nordlander, }1998]{Nordlander98}
Nordlander, J. (1998).
\newblock {Pragmatic Subtyping in Polymorphic Languages}.
\newblock {\em Pages  216--227 of:} Berman, M., \& Berman, S. (eds), {\em
  Proceedings of the third {ACM} {SIGPLAN} international conference on
  functional programming ({ICFP}-98)}.
\newblock ACM SIGPLAN Notices, vol. 34, 1.
\newblock New York: ACM Press.

\bibitem[\protect\citename{Nordlander, }2002]{Nordlander02}
Nordlander, J. (2002).
\newblock Polymorphic subtyping in {O'Haskell}.
\newblock {\em Science of computer programming}, {\bf 43}(2--3), 93--127.
\newblock Also in the Proceedings of the APPSEM Workshop on Subtyping and
  Dependent Types in Programming, Ponte de Lima, Portugal, 2000.

\bibitem[\protect\citename{Ohori, }1995]{Ohori95}
Ohori, A. (1995).
\newblock {A polymorphic record calculus and its compilation}.
\newblock {\em {TOPLAS}}, {\bf 17}(6), 844--895.

\bibitem[\protect\citename{Pang \& Chakravarty, }2004]{PC04}
Pang, A.T.H., \& Chakravarty, M.M.T. (2004).
\newblock {Interfacing Haskell with Object-Oriented Languages}.
\newblock {\em Pages  20--35 of:} {P.W.~Trinder and G.~Michaelson and R.~Pena}
  (ed), {\em {Implementation of Functional Languages, 15th International
  Workshop, IFL 2003, Edinburgh, UK, September 8-11, 2003, Revised Papers}}.
\newblock LNCS, vol. 3145.
\newblock Springer-Verlag.

\bibitem[\protect\citename{{Peyton Jones}, }2001]{spj00}
{Peyton Jones}, S.L. (2001).
\newblock {Tackling the Awkward Squad: monadic input/output, concurrency,
  exceptions, and foreign-language calls in {Haskell}}.
\newblock {\em Pages  47--96 of:} Hoare, C.A.R., Broy, M., \& Steinbrueggen, R.
  (eds), {\em {Engineering theories of software construction, {Marktoberdorf
  Summer School} 2000}}.
\newblock NATO ASI Series.
\newblock IOS Press.

\bibitem[\protect\citename{{Peyton Jones} \& Wadler,
  }1993]{peytonjoneswadler-popl93}
{Peyton Jones}, S.L., \& Wadler, P. 1993 (Jan.).
\newblock Imperative functional programming.
\newblock {\em Pages  71--84 of:} {\em the symposium on principles of
  programming languages ({POPL} '93)}.
\newblock ACM.

\bibitem[\protect\citename{{Peyton~Jones} {\em et~al.}\relax, }1997]{PJJM97}
{Peyton~Jones}, S.L., Jones, M.P., \& Meijer, E. (1997).
\newblock Type classes: exploring the design space.
\newblock  Launchbury, J. (ed), {\em {Haskell Workshop}}.

\bibitem[\protect\citename{Pierce \& Turner, }1994]{PT94}
Pierce, B.C., \& Turner, D.N. (1994).
\newblock {Simple Type-Theoretic Foundations for Object-Oriented Programming}.
\newblock {\em Journal of functional programming}, {\bf 4}(2), 207--247.

\bibitem[\protect\citename{R{\'e}my, }1994]{ML-ART}
R{\'e}my, D. (1994).
\newblock {Programming Objects with {ML-ART}: An extension to {ML} with
  Abstract and Record Types}.
\newblock {\em Pages  321--346 of:} Hagiya, M., \& Mitchell, J.C. (eds), {\em
  {International Symposium on Theoretical Aspects of Computer Software}}.
\newblock LNCS, no.  789.
\newblock Sendai, Japan: Springer-Verlag.

\bibitem[\protect\citename{R{\'e}my \& Vouillon, }1997]{RV97}
R{\'e}my, D., \& Vouillon, J. (1997).
\newblock {Objective ML: a simple object-oriented extension of ML}.
\newblock {\em Pages  40--53 of:} {\em {POPL '97: Proceedings of the 24th ACM
  SIGPLAN-SIGACT Symposium on Principles of Programming Languages}}.
\newblock New York, NY, USA: ACM Press.

\bibitem[\protect\citename{Shields \& {Peyton Jones}, }2001]{SPJ01}
Shields, M., \& {Peyton Jones}, S.L. (2001).
\newblock {Object-Oriented Style Overloading for Haskell}.
\newblock {\em {ENTCS}}, {\bf 59}(1).
\newblock An extended TR is available from
  \url{http://research.microsoft.com/Users/simonpj/Papers/oo-haskell/}.

\bibitem[\protect\citename{Siek {\em et~al.}\relax,
  }2005]{siek05:_concepts_cpp0x}
Siek, J., Gregor, D., Garcia, R., Willcock, J., J\"arvi, J., \& Lumsdaine, A.
  (2005).
\newblock {\em Concepts for {C++0x}}.
\newblock Tech. rept. N1758=05-0018. ISO/IEC JTC 1, Information Technology,
  Subcommittee SC 22, Programming Language {C++}.

\bibitem[\protect\citename{Stuckey \& Sulzmann, }2005]{SS05}
Stuckey, P.~J., \& Sulzmann, M. (2005).
\newblock {A Theory of Overloading}.
\newblock {\em {TOPLAS}}.
\newblock To appear.

\bibitem[\protect\citename{Stuckey {\em et~al.}\relax, }2004]{SSW04}
Stuckey, P.~J., Sulzmann, M., \& Wazny, J. (2004).
\newblock Improving type error diagnosis.
\newblock {\em Pages  80--91 of:} {\em Proc.\ of haskell'04}.
\newblock ACM Press.

\bibitem[\protect\citename{Surazhsky \& Gil, }2004]{SG04}
Surazhsky, V., \& Gil, J.Y. (2004).
\newblock {Type-safe covariance in {C++}}.
\newblock {\em Pages  1496--1502 of:} {\em {SAC '04: Proceedings of the 2004
  ACM Symposium on Applied Somputing}}.
\newblock ACM Press.

\bibitem[\protect\citename{Taha \& Nielsen, }2003]{env-classifiers}
Taha, W., \& Nielsen, M.F. (2003).
\newblock {Environment classifiers}.
\newblock {\em Pages  26--37 of:} {\em {POPL '03: Proceedings of the 30th ACM
  SIGPLAN-SIGACT Symposium on Principles of Programming Languages}}.
\newblock New York, NY, USA: ACM Press.

\bibitem[\protect\citename{Ungar \& Smith, }1987]{Self}
Ungar, D., \& Smith, R.~B. (1987).
\newblock {Self: The power of simplicity}.
\newblock {\em Pages  227--242 of:} {\em {OOPSLA 1987: Conference Proceedings
  on Object-Oriented Programming Systems, Languages and Applications}}.
\newblock ACM Press.

\bibitem[\protect\citename{Zenger \& Odersky, }2004]{ZO04}
Zenger, M., \& Odersky, M. 2004 (Mar.).
\newblock {\em {Independently Extensible Solutions to the Expression Problem}}.
\newblock Tech. rept. Ecole Polytechnique Féd{\'e}rale de Lausanne.
\newblock Technical Report Nr. 200433.

\end{thebibliography}

}


\end{document}